\documentclass[apj]{emulateapj}

\slugcomment{}

\shorttitle{Dust Reverberation of Seyfert Galaxies}
\shortauthors{Koshida et al.}

\begin{document}

\title{Reverberation Measurements of the Inner Radius of the Dust Torus in 17 Seyfert Galaxies}

\author{Shintaro Koshida\altaffilmark{1},
Takeo Minezaki\altaffilmark{2}, Yuzuru Yoshii\altaffilmark{2,3},
Yukiyasu Kobayashi\altaffilmark{4}, Yu Sakata\altaffilmark{2,5},
Shota Sugawara\altaffilmark{2,5}, Keigo Enya\altaffilmark{6},
Masahiro Suganuma\altaffilmark{4}, Hiroyuki Tomita\altaffilmark{7},
Tsutomu Aoki\altaffilmark{7}, and Bruce A. Peterson\altaffilmark{8}}

\altaffiltext{1}{Center of Astro Engineering and Department of Electrical Engineering,
Pontificia Univercsidad Catolica de Chile, Av. Vicuna Mackenna 4868,
Chile; skoshida@ing.puc.cl}
\altaffiltext{2}{Institute of Astronomy, School of Science, University
of Tokyo, 2-21-1 Osawa, Mitaka, Tokyo 181-0015, Japan; minezaki@ioa.s.u-tokyo.ac.jp}
\altaffiltext{3}{Research Center for the Early Universe, School of Science, University of Tokyo, 7-3-1 Hongo, Bunkyo-ku, Tokyo 113-0013, Japan}
\altaffiltext{4}{National Astronomical Observatory, 2-21-1 Osawa, Mitaka, Tokyo 181-8588, Japan}
\altaffiltext{5}{Department of Astronomy, School of Science, University of Tokyo, 7-3-1 Hongo, Bunkyo-ku, Tokyo 113-0013, Japan}
\altaffiltext{6}{Institute of Space and Astronautical Science, Japan Aerospace Exploration Agency, 3-1-1, Yoshinodai, Sagamihara, Kanagawa 229-8510, Japan}
\altaffiltext{7}{Kiso Observatory, Institute of Astronomy, School of Science, University of Tokyo, 10762-30 Mitake, Kiso, Nagano 397-0101, Japan}
\altaffiltext{8}{Mount Stromlo Observatory, Research School of Astronomy and Astrophysics, Australian National University, Weston Creek P.O., ACT 2611, Australia}

\begin{abstract}
We present the results of a dust reverberation survey
for 17 nearby Seyfert 1 galaxies, which provides the largest
homogeneous data collection for the radius of the innermost dust torus.
A delayed response of the $K$-band light curve
after the $V$-band light curve was found for all targets,
and 49 measurements of lag times between
the flux variation of the dust emission in the $K$ band
and that of the optical continuum emission in the $V$ band
were obtained by the cross-correlation function analysis
and also by an alternative method for estimating the maximum likelihood lag.
The lag times strongly correlated with the optical luminosity
in the luminosity range of $M_{V}=-16$ to $-22$ mag,
and the regression analysis was performed to obtain the correlation
$\log \Delta t\ ({\rm days})\ = -2.11 -0.2M_{V}$
assuming $\Delta t \propto L^{0.5}$, which was theoretically expected.
We discuss the possible origins of the intrinsic scatter
of the dust lag-luminosity correlation, which was estimated
to be approximately $0.13$ dex,
and we find that the difference of internal extinction
and delayed response of changes in lag times
to the flux variations could have partly contributed
to intrinsic scatter.
However, we could not detect any systematic change of
the correlation with the subclass of the Seyfert type or the Eddington ratio.
Finally, we compare the dust reverberation radius with
the near-infrared interferometric radius of the dust torus
and the reverberation radius of broad Balmer emission lines.
The interferometric radius in the $K$ band was found to be
systematically larger than the dust reverberation radius in the same band
by about a factor of two, which could be interpreted by the difference
between the flux-weighted radius and the response-weighted radius
of the innermost dust torus.
The reverberation radius of the broad Balmer emission lines
was found to be systematically smaller than the dust reverberation
radius by about a factor of $4$--$5$,
which strongly supports the unified scheme of the Seyfert type of
active galactic nuclei (AGNs).
Moreover, we examined the radius-luminosity correlations
for the hard X-ray ($14$--$195$ keV) and
the [\ion{O}{4}]$\lambda 25.89$ $\mu$m emission-line luminosities,
which would be applicable for obscured AGNs.

\end{abstract}


\section{Introduction}

An active galactic nucleus (AGN) is known as one of the most luminous
populations in the Universe, and its enormous radiative energy is powered
by mass accretion onto the supermassive black hole (SMBH).
Many observations have indicated that
an optically and geometrically thick torus consisting of gas and dust
surrounds the SMBH and accretion disk
\citep[e.g.,][]{tele+84,anto+85,mill+90},
which could be a gas reservoir for supplying accreting mass onto the
accretion disk \citep[e.g.,][]{krolik+88}.
This dust torus is regarded as a key structure for the unified scheme of
AGNs, which explains differences between type-1 and type-2 AGNs
with respect to the viewing angle and obscuration by the dust torus
\citep[e.g.,][]{anto93}.
In this scheme, type-1 AGNs with broad emission lines (BELs) are
classified as type-2 AGNs if the broad emission-line region (BLR) is
obscured by the dust torus.

The nature of the dust torus has been extensively investigated
by observational and theoretical methods.
The near-infrared continuum emission of type-1 AGNs is
considered to be dominated by the thermal re-radiation of hot dust
driven by the reprocessing
of the ultraviolet (UV)--optical continuum emission from the accretion disk
\citep[e.g.,][]{rees+69, barv87, koba+93, land+11}.
In addition, significant research has been conducted to explain
the infrared spectral energy distribution (SED) of AGNs
and to understand the dust torus structure
\citep[e.g.,][]{pk92, nenk+02, nenk+08, dull+05,
schartmann+05, schartmann+08, honi+06, mor+09, deo+11}.
However, since the apparent size of the dust torus is so compact,
it has been almost impossible to resolve and examine 
the detailed structure of the dust torus directly
by imaging observation.

Reverberation mapping observation 
provides a unique and important tool for investigating
the structures of innermost dust tori
\citep{clav+89, bari+92, glas92, sitk+93,
okny93, nels96a, okny+99, nels+01, okny+01,glas04,lira+11,pozo+14}.
A time lag between the variation of optical flux
originating in the accretion disk
and that of near-infrared flux in the innermost dust torus
can be interpreted as the light-travel time
from the accretion disk to the dust torus,
which corresponds to the radius of the innermost dust torus.
The Multicolor Active Galactic Nuclei Monitoring (MAGNUM) project
(Yoshii 2002; Yoshii, Kobayashi, \& Minezaki 2003)
conducted extensive monitoring observations in
optical and near-infrared wavelengths for a number of type-1 AGNs to
obtain precise estimates of the radius of the innermost dust torus.
Indeed, lag times of weeks to months corresponding to $0.01$--$0.1$ pc
were obtained for several Seyfert galaxies from
optical and near-infrared monitoring observations,
in which a strong correlation
between lag time and optical luminosity was found to be
consistent with $\Delta t_{\rm dust}\propto L^{0.5}$
\citep{mine+04,suga+04,suga+06,kosh+09}.
Moreover, \citet{suga+06} reported that the dust lag placed
an upper boundary on the lag of the BELs,
which strongly supports the unified scheme of AGNs.

Recent advancement in infrared interferometry has enabled
milliarcsecond-scale measurement of the dust tori
for bright AGNs, which has become an important tool for studying
the structure of the dust torus.
Near-infrared interferometric observations have been
applied to less than 10 type-1 AGNs for estimating
the radius of the innermost dust torus
\citep{swai+03, kish+09, pott+10, kish+11, weigelt+12}.
These radii were found to be scaled approximately
as $r_{\rm dust}\propto L^{0.5}$,
and their values were found to be of the same order
as the dust reverberation radii
\citep{kish+09, kish+11}.

Such luminosity correlation is expected from a model in which
the dust temperature and the inner radius of the dust torus are
determined by radiation equilibrium and sublimation of dust, respectively.
\citet{barv87} quantitatively estimated the inner radius of
the dust torus by considering the wavelength-dependent
absorption efficiency of dust grains,
and \citet{kish+07} expressed
dependency on dust grain size as
\begin{equation}
\footnotesize
  R_{\rm{sub}}=1.3\left (\frac{L_{\rm {UV}}}{10^{46}{\rm { erg/s}}}\right )^{0.5}\left(\frac{T_{\rm {sub}}}{1500\rm{K}}\right)^{-2.8}\left(\frac{a}{0.05 \mu\rm{m}}\right)^{-0.5}~\rm{pc},\label{eq:intro:1}
\end{equation}
where $L_{\rm {UV}}$, $T_{\rm {sub}}$, and $a$ are
the UV luminosity of the accretion disk,
sublimation temperature of dust,
and dust grain size, respectively.
Although the previously observed luminosity dependency
of the radius of the innermost dust torus
is consistent with the theoretical expectation,
\citet{kish+07} found that the innermost torus radii based on
dust reverberation were systematically smaller than
those based on the theoretical prediction of Equation (1),
assuming $T_{\rm sub}=1500$ K and $a=0.05\ \mu$m, by a factor of $\sim 3$.
They proposed that this discrepancy could be explained by a higher
sublimation temperature or a larger grain size, although the former case
would be disfavored by the near-infrared colors of AGNs.
On the contrary, \citet{kawa+10,kawa+11} developed
a reverberation model for a clumpy dust torus
considering the anisotropic illumination of the accretion disk.
They presented that dust clouds could survive
closer to the accretion disk on the equatorial plane
than at the expected sublimation radius by assuming isotropic illumination,
which could explain the dust lag discrepancy.
They further reported that according to the model,
the dust lag increases with the viewing angle of the dust torus,
and they suggested that the difference in the viewing angle
is primarily responsible for the scatter of
the dust lag-luminosity correlation.

In this study, we present 49 measurements of lag time between
optical and near-infrared light curves for 17 nearby AGNs
obtained by the MAGNUM project,
which is the largest collection of
systematic survey data of dust reverberation.
By estimating the radii of the innermost dust tori from the data,
we discuss the structure of AGN dust tori.
In Section \ref{sec:obs}, we describe the target AGNs and
the procedures of observation, reduction, and photometry.
In addition, we describe the subtraction of contaminated flux
by the host-galaxy starlight and narrow emission lines
for estimating the optical continuum emission from the accretion disk.
In Section \ref{sec:ana},
we measure the lag time between optical and near-infrared
flux variations by using cross-correlation function (CCF) analysis.
In Section \ref{sec:laglum}, we examine the lag-luminosity correlation
and its scatter on the basis of our uniformly analyzed
largest dust reverberation sample.
In Section \ref{sec:dis},
we compare the structure around the BLR and the innermost dust torus
determined by our reverberation results
with those of near-infrared interferometry and
BEL reverberation.
In addition, we discuss the secondary dependence
of the dust lag on the Seyfert subclass and the mass accretion rate.
We summarize the results in Section \ref{sec:sum}.
We assume the cosmology of $H_{0}=73$ km s$^{-1}$ Mpc$^{-1}$, $\Omega_{m}=0.27$, and
$\Omega_{\Lambda}=0.73$ according to \citet{sper+07}
throughout this study.

\section{Observations and Photometry}\label{sec:obs}

We briefly describe the procedures of observation, data reduction,
photometry, and estimation of host-galaxy and narrow emission-line
fluxes.
These procedures follow those presented in previous studies
\citep{suga+06, kosh+09, saka+10}.

\subsection{Targets}

\begin{figure}
\epsscale{0.9}
\plotone{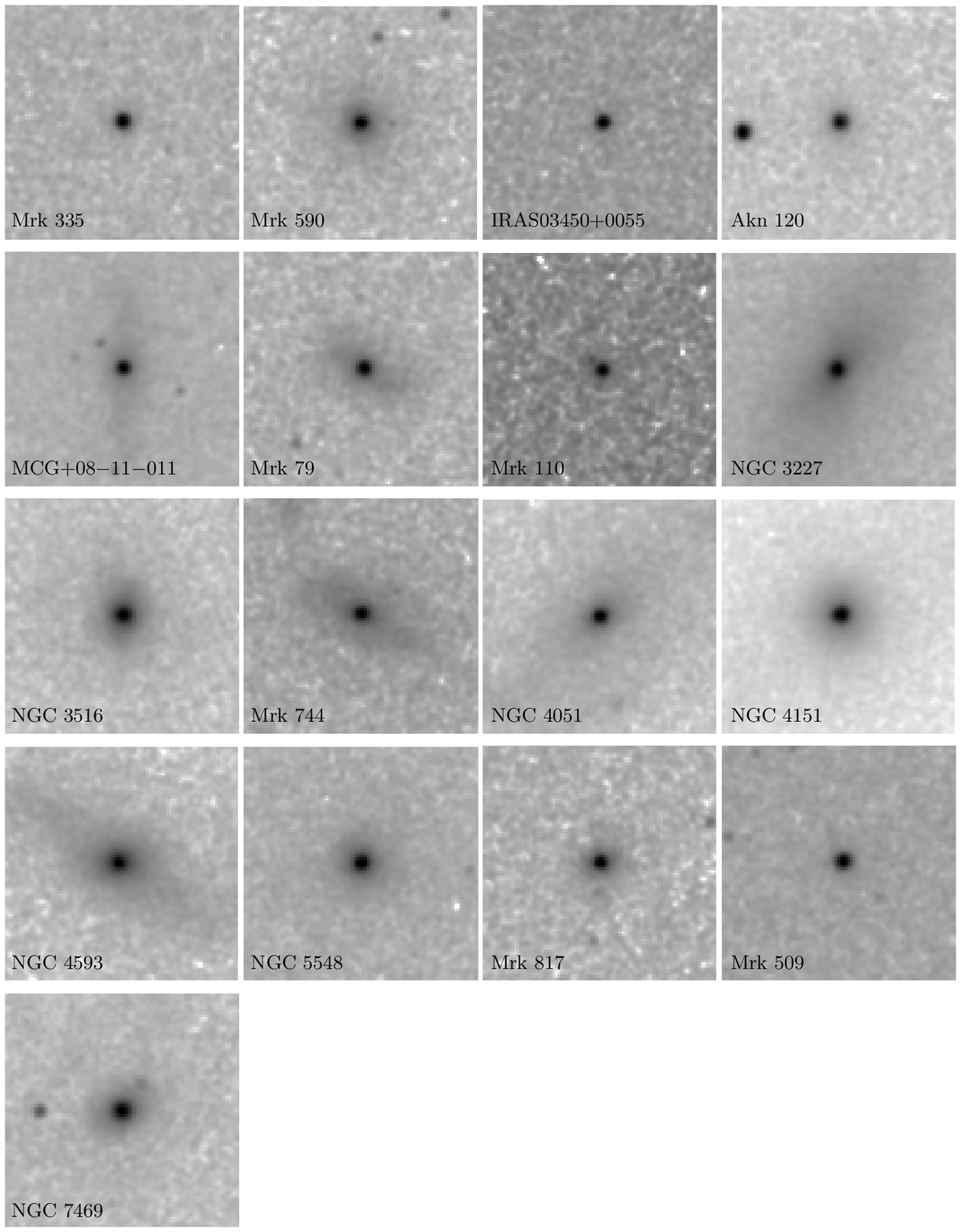}
\caption[The Portrait images of the target Seyfert galaxies]{
Portrait images of the target Seyfert galaxies in the $K$ band
(taken from the Two Micron All Sky Survey).
The field of fiew of the image is $1.5\times 1.5$ arcmin$^2$,
which represents that of the MAGNUM MIP camera for this study.
The image intensity levels are displayed in logarithmic scale. 
}\label{fig:portrait}
\end{figure}

We focus on $17$ type-1 AGNs from the MAGNUM targets
that exist in the local Universe with redshifts of less than 0.05.
The target AGNs and their basic parameters are listed in Table
\ref{tab:obs:tar:stdinfo},
and their portrait images are presented in Figure \ref{fig:portrait}.
Because these nearby Seyfert galaxies are less luminous,
their timescales of flux variation are considered to be
relatively short, and
multiple features such as local peaks and valleys in the light curve
enable the precise measurement of the lag times.
In addition, their host-galaxy flux can be estimated by
surface brightness fitting more reliably than
that of distant quasi-stellar objects (QSOs).
Of the 17 target AGNs, 14 were selected from the type-1 AGNs for which
the BEL lag had been determined by reverberation mapping observations
\citep{wpm99,onke+03,denn+06};
two targets, IRAS 03450$+$0055 and Mrk 744, were selected from those
for which dust reverberation mapping observations had been 
performed \citep{nels96a, nels96b};
and the last target, MCG $+$08-11-011, was taken from \citet{enya+02}
for which large amplitudes of flux variation in near-infrared wavebands
had been detected.
The optical luminosity of the targets ranged from $M_{V}= -15.8$ to
$-22.2$ after the host-galaxy flux was subtracted (see Section
\ref{subsec:host}).

\subsection{Observations}\label{subsec:obs}

Monitoring observations were conducted by using the multicolor imaging
photometer (MIP) mounted on the MAGNUM telescope
\citep{koba+98a,koba+98b}. 
The MIP has a field-of-view of 1.5$\times$1.5 arcmin$^{2}$;
it is capable of simultaneously obtaining images
in optical ($U$, $B$, $V$, $R$, and $I$) and
near-infrared ($J$, $H$, and $K$) bands by splitting the incident beam into
two different detectors including an SITe CCD (1024$\times$1024 pixels, 0.277
arcsec pixel$^{-1}$) and an SBRC InSb array (256$\times$256 pixels,
0.346 arcsec pixel$^{-1}$).

Observations were conducted most frequently in the $V$ and $K$ bands.
The $V$ band aims to obtain optical continuum emission from the accretion
disk, avoiding the contamination of variable BELs such as the Balmer series.
On the contrary, the $K$ band aims to obtain reprocessed thermal emission
of hot dust located at the innermost region of the dust torus, 
in which such thermal emission dominates over
the power-law continuum emission from the accretion disk.
In this study, we focus on the results derived from the light curves in
these bands.

Monitoring observations with the MAGNUM telescope began
in 2001--2003, although that for NGC 3516 and NGC 4593
began in 2005.
We present the data obtained through 2006--2007
to include monitoring spans of 3--7 years.
The typical monitoring intervals were configured to be shorter
for less luminous targets for which shorter lag times were expected,
and they were from a few days to 20 days.
The observational parameters are listed in Table \ref{tab:obs:sch}.

\subsection{Reduction and Photometry}\label{subsec:red}

The images were reduced using IRAF 
\footnote{IRAF is distributed by the National Optical Astronomy Observatories, which are operated by the Association of Universities for Research in Astronomy, Inc., under cooperative agreement with the National Science Foundation.}.
We followed the standard procedures for image reduction such as bias or
sky subtraction and flat fielding, with small corrections applied
for non-linear detector response. 

\begin{figure}
\epsscale{0.7}
\plotone{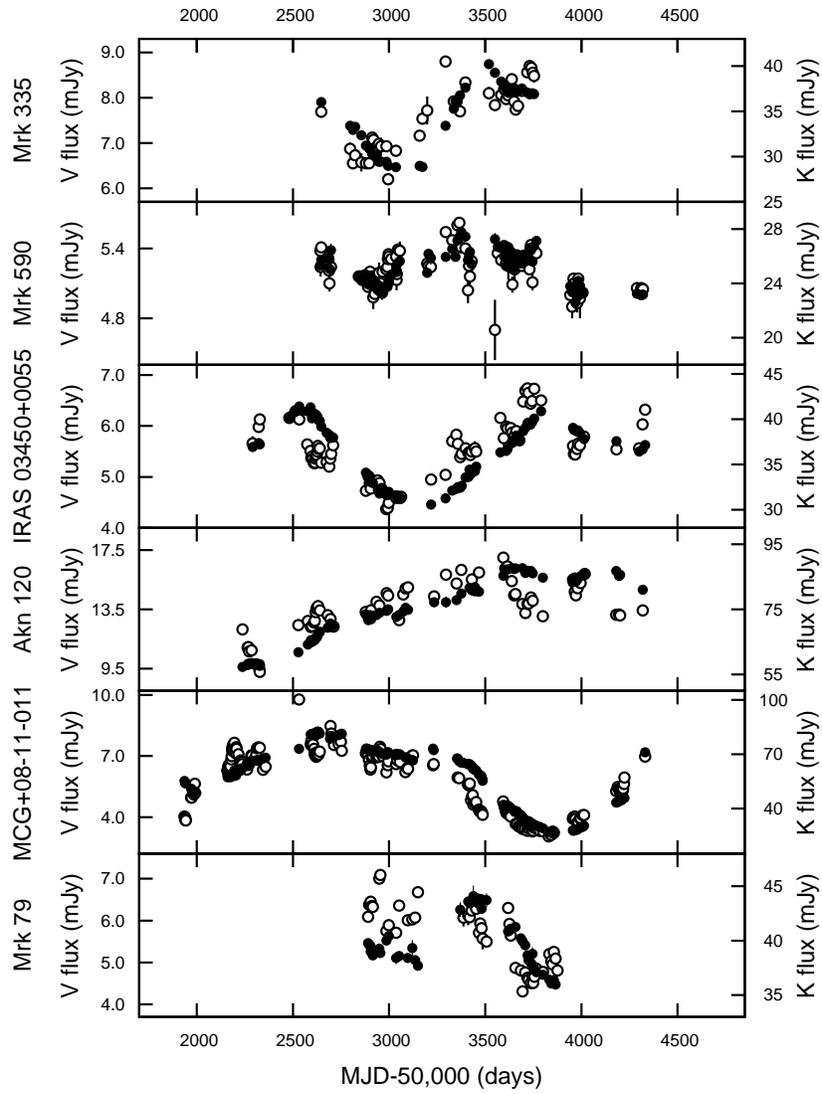}
\caption[The light curves of the targets]{
$V$-band (open circles) and $K$-band (filled circles) light
curves of Mrk 335, Mrk 590, IRAS 03450$+$0055, Akn 120, MCG $+$08-11-011,
and Mrk 79.
Correction for the Galactic extinction
has not been applied to the light curves,
and the fluxes from the host galaxy and narrow emission lines
have not been subtracted.
}\label{fig:lc1}
\end{figure}

\begin{figure}
\epsscale{0.7}
\plotone{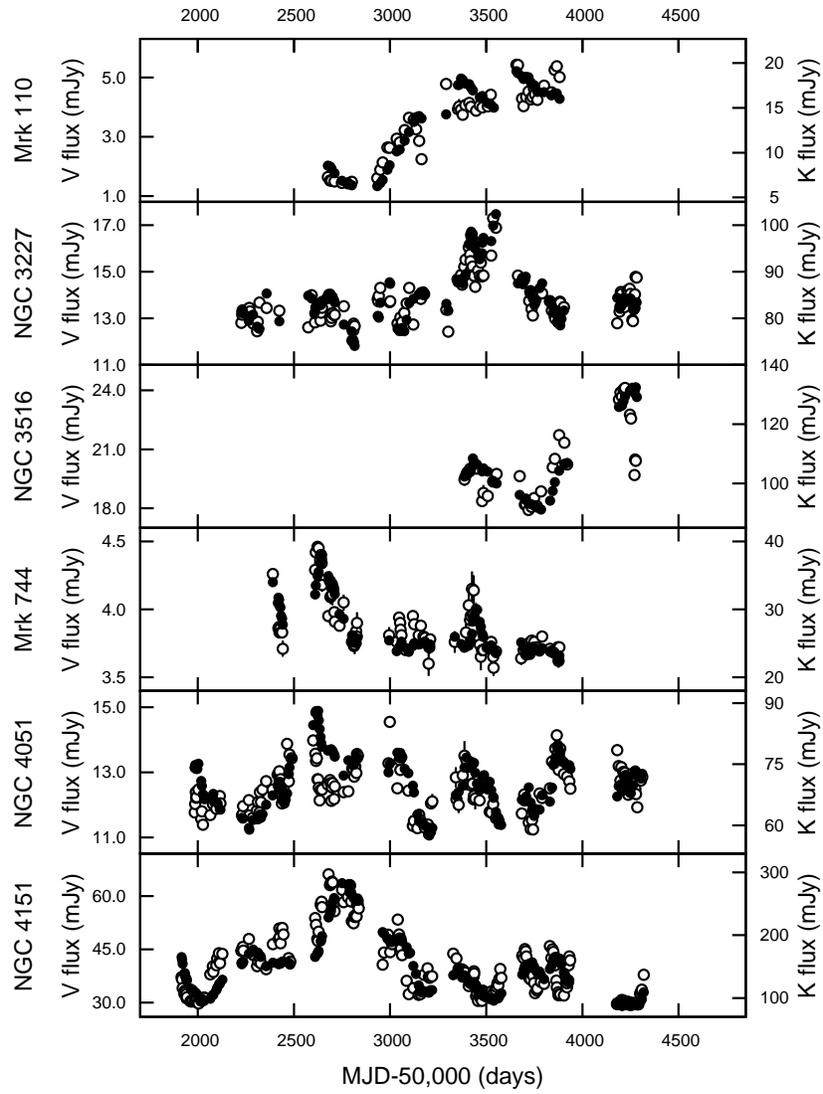}
\caption[The light curves of the targets (Continued)]{
$V$-band and $K$-band light curves of
Mrk 110, NGC 3227, NGC 3516, Mrk 744, NGC 4051, and NGC 4151.
Symbols and others are the same as Figure \ref{fig:lc1}.
}\label{fig:lc2}
\end{figure}

\begin{figure}
\epsscale{0.7}
\plotone{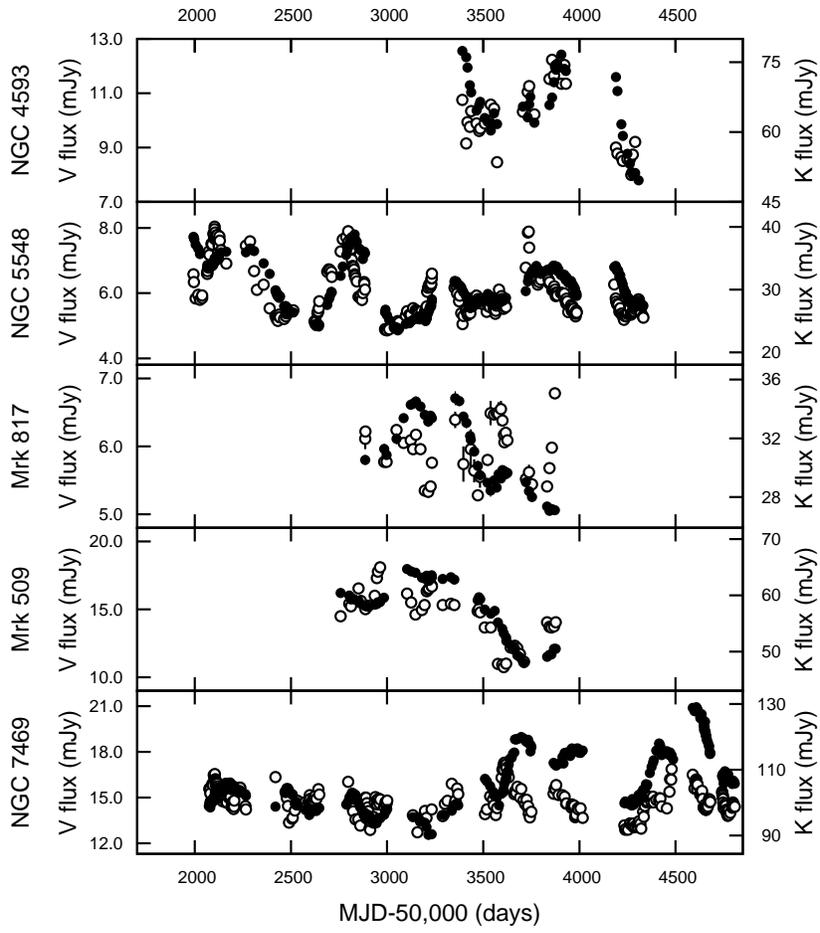}
\caption[The light curves of the targets (Continued)]{
$V$-band and $K$-band light curves of
NGC 4593, NGC 5548, Mrk 817, Mrk 509, and NGC 7469.
Symbols and others are the same as Figure \ref{fig:lc1}.
}\label{fig:lc3}
\end{figure}

As shown in Figure \ref{fig:portrait},
the MIP's limited field-of-view prevented us from obtaining
suitable stars for the reference of the point-spread function (PSF) shape
and photometry in the same images as those of bright and extended targets.
Therefore, we observed the target AGN and its reference stars alternately
to measure the nuclear flux of the target with respect to
these reference stars,
and aperture photometry
with a circular aperture of $\phi = 8\arcsec.3$ in diameter
with a sky reference area of a $\phi = 11\arcsec.1-13\arcsec.9$
annulus was applied
instead of difference imaging photometries
\footnote{
We also note that the PSFs of the target image and the reference star image
are slightly different because they were not simultaneously obtained,
which makes it difficult to employ difference imaging photometries.}.
Although a relatively large aperture was selected to keep the photometry
stable against the seeing variation,
the photometric error of the nuclear flux relative to the reference
stars was typically $0.01$ mag.
The fluxes of the reference stars were calibrated with respect to
photometric standard stars designated by \citet{land92} and \citet{hunt+98}
for the $V$ and $K$ bands, respectively, and the errors in flux
calibration were typically less than $0.01$ mag.
The positions and magnitudes of the reference stars are listed in
Table \ref{tab:red:phot:abs:refstar1}.
The resultant light curves of the 17 target AGNs in the $V$ and $K$ bands
are presented in Figures \ref{fig:lc1}--\ref{fig:lc3},
and their data are listed in Table \ref{tab:lc}.

\subsection{Subtraction of Host-Galaxy Flux}\label{subsec:host}
The flux obtained by aperture photometry
contained a significant contribution of host-galaxy starlight
and thus was subtracted.
Nine of the 17 target AGNs are common to those of \citet{saka+10}, who
estimated the host-galaxy fluxes in the optical $B$, $V$, and $I$ bands
within the same photometric aperture by surface brightness fitting to
the high-resolution {\it Hubble Space Telescope (HST)}
Advanced Camera for Surveys images
as well as those obtained by MAGNUM. 
We adopted the same $V$-band fluxes of host galaxies reported by
\citet{saka+10} for these targets.
In addition, we adopted the results by \citet{suga+06},
who estimated the host-galaxy flux of NGC 7469 in
the $V$ band within the same photometric aperture.

For the remaining seven target AGNs,
we estimated the host-galaxy flux in the $V$ band
by surface brightness fitting in the same manner
as that reported by \citet{suga+06}.
We first selected the data obtained at night
with good and stable seeing conditions,
typically with a PSF full-width at half-maximum of less than 1.5 arcsec.
All target AGN and reference-star images obtained
during the same night were then combined into one image.
The target AGN image was fitted by the GALFIT two-dimensional image
decomposition program \citep{peng+02} with analytic functions
for the galaxy component of bulge (de Vaucouleurs' R$^{1/4}$ profile) and
disk (exponential profile), plus a PSF for the AGN nucleus,
using the images of the reference stars simultaneously observed
as the PSF reference.
Finally, aperture photometry was applied to the nucleus-free image
created by subtracting the best-fit PSF component from the original image.
We adopted the average and scatter of the aperture fluxes of the
nuclear-free images over the data obtained during various nights
as the host-galaxy flux and its error.

For the host-galaxy fluxes in the $K$ band, we adopted the results of
\citet{mine+04} for NGC4151; those of \citet{suga+06} for
NGC 5548, NGC 4051, NGC 3227, and NGC 7469;
those of \citet{tomi+06} for MCG $+$08-11-011;
and those of \citet{tomi05} for IRAS 03450$+$0055, Akn 120.
For the remaining 9 target AGNs,
we estimated the host-galaxy flux in the $K$ band
by surface brightness fitting in the same manner as
that conducted for the $V$ band.
The host-galaxy fluxes within the aperture and their errors are listed
in Table \ref{tab:hostflux}.

\subsection{Subtraction of Narrow Emission-Line Flux}\label{subsec:nl}
In addition to subtracting the contaminated flux of host-galaxy starlight,
we subtracted the contribution of the two strong narrow emission lines,
[\ion{O}{3}]$\lambda 4959$ and $\lambda5007$,
from the $V$-band flux for estimating the flux of the optical
continuum emission from the accretion disk.
\citet{saka+10} estimated the contribution of the narrow emission
lines in the $V$ band; therefore, we adopted their estimates for the nine common targets.

For the remaining eight target AGNs, we estimated the contribution of
the narrow emission lines in the $V$ band in the same manner as
that conducted by \citet{saka+10}.
The flux of [\ion{O}{3}]$\lambda5007$ was compiled from previous research
\citep{boro+92,whit92,kasp+96,marz+03,schm+03,kasp+05}, and the flux of
[\ion{O}{3}]$\lambda4959$ was calculated by assuming a theoretical line
ratio of [\ion{O}{3}]$\lambda5007/$[\ion{O}{3}]$\lambda4959=3.01$
\citep{stor+00}. 
We then estimated their flux contributions by convolving
a narrow-line spectrum with the transmission curve of the $V$-band filter.
Following \citet{saka+10}, we adopted $10$\% of the flux of
the narrow-line component as its error.
Contributions of the narrow emission lines are also
listed in Table \ref{tab:hostflux}.

\section{Dust Reverberation Analysis}\label{sec:ana}
As presented in Figures \ref{fig:lc1}--\ref{fig:lc3},
all targets showed significant flux variations in both bands
and delayed responses of the $K$-band light
curve after that of the $V$ band, which indicate that
the reprocessed thermal radiation of hot dust
dominated in the $K$-band flux.

In this section, we estimate the lag time
between the $V$- and $K$-band light curves,
which corresponds to the radius of the innermost dust torus,
by using CCF analysis \citep[][]{whit+94,pete01}
as well as by an alternative method for
estimating the maximum likelihood lag that was
recently developed by Zu, Kochanek, \& Peterson (2011).
Following the method of \citet[][]{kosh+09},
we first subtracted the accretion-disk component of flux
in the $K$ band to derive the dust torus component,
then we applied the CCF analysis
and the alternative method
to the $V$-band light curve
and that of the dust torus component in the $K$ band.
In this study, we summarize only the procedures;
details for the subtraction of
the accretion-disk component and the CCF analysis
have been reported in previous studies \citep{suga+06,kosh+09},
and those for the alternative method are described in \citet[][]{zu+11}.

\subsection{Subtraction of the Accretion-Disk Component from the K-Band Flux}\label{subsec:ana:lc}

Although the thermal radiation of the dust torus usually dominates in
the $K$-band flux of type-1 AGNs, it contains emission from
the accretion disk \citep{tomi+06,kish+08,lira+11}. 
While the host-galaxy flux has no influence on CCF analysis
because it is not variable, the superimposed flux variation by
the accretion-disk component in the $K$-band flux
would make the lag time obtained by CCF analysis
shorter than the actual lag of the dust-torus emission.
Such a result occurs, because the power-law continuum emission
from the accretion disk in the near-infrared spectra varies with time
and is nearly synchronous with that in the optical
\citep{tomi+06,mine+06,lira+11}.

Therefore, we estimated the contribution of the accretion disk in the
$K$-band flux at a certain epoch, $f_{K,\rm{disk}}(t)$,
and subtracted it from the observed $K$-band flux prior to CCF analysis.
In this study, $f_{K,\rm{disk}}(t)$ was estimated from the $V$-band flux
by assuming a power-law spectrum of the accretion-disk emission as
\begin{equation}
f_{K,\rm{disk}}(t)=f_V(t) \left(\frac{\nu_K}{\nu_V}\right)^{\alpha_\nu},
\end{equation}
where $f_V(t)$ is the $V$-band flux at the same epoch after
the host-galaxy and narrow emission-line fluxes were subtracted;
$\nu_K$ and $\nu_V$ are effective frequencies
of the $K$ and $V$ bands, respectively;
and $\alpha_\nu$ is the power-law index.
If no optical photometric data were available at the same epoch,
$f_V(t)$ was obtained by the linear interpolation
of the photometric data observed at the closest epochs.
Prior to the estimation and subsequent subtraction of
the accretion-disk component, $f_{K,\rm{disk}}$,
the observed $V$- and $K$-band fluxes were corrected for
the Galactic extinction according
to the NASA/IPAC Extragalactic Database
\citep[NED; based on][]{schl+98}.

The spectral shape of the flux variation
of accretion-disk emission in optical and
near-infrared wavelengths has not been well established.
\citet{tomi+06} applied the multiple regression analysis 
to the flux variations observed in optical and near-infrared
wavebands of a Seyfert galaxy MCG $+$08--11--011
to estimate the power-law index of the accretion-disk
component of the flux variation in optical and near-infrared spectra
as $\alpha_\nu \sim -0.1$ to $+0.4$.
From the same analysis applied to a number of nearby Seyfert
galaxies by \citet{tomi05}, the value could be estimated 
as $\alpha_\nu =+0.10\pm 0.11$.
According to the standard accretion model \citep{shak+73},
the power-law index $\alpha_\nu $ increases with wavelength
and converges asymptotically on $\alpha_\nu = +1/3$
in the limit of long wavelength,
and the power-law index of the flux variation would be
the same in the long-wavelength limit
when the standard accretion model is retained
during flux variations.
\citet{wilh+05} derived the composite differential
spectrum of QSOs observed by the Sloan Digital Sky Survey
and found that it could be fitted well
by the power-law form of $\alpha_\nu =0$,
although the wavelength range was UV to optical.
\citet{kish+08} conducted spectropolarimetry for
radio-loud and -quiet QSOs and obtained $\alpha_\nu =0.44\pm 0.11$,
although the observation was single epoch.
From these considerations, we calculated $f_{K,\rm{disk}}$
assuming both $\alpha_\nu = 0$ and $+1/3$
to examine a systematic difference
in the dust lag estimation.

\subsection{Measuring the Dust Lag by the CCF Analysis}

After subtracting the accretion-disk component from the $K$-band flux,
we measured the lag time between the flux variation
of the thermal emission from the dust torus in the $K$ band
and that of the optical continuum emission from the accretion disk
in the $V$ band.
First, we applied widely used, conventional CCF methods
that have been employed
 \citep{whit+94,pete01,raft+11,bart+11a,bart+11b,doro+12,edri+12,grie+12a,2014ApJ...782...45D}.
For comparison with our new results,
the bi-directional interpolation (BI) method was adopted,
which is the same method as that used in our previous studies
\citep{suga+06,kosh+09}.
The BI method is used to calculate the CCF
of the light curves with irregular sampling obtained
by monitoring observations,
and it incorporates interpolation errors caused by
the target's flux variation, as is described subsequently.

With an arbitrary shift of time between the light curves
in two different bands, $\tau$, we generated the data pairs of fluxes
whereby the flux in one band was obtained by actual observation
and that in the other band was estimated by the linear interpolation
of the light curve.
Once the data pairs were produced for all observed data of both bands,
a cross correlation for $\tau$ was calculated
by using the set of data pairs.
The CCF could then be calculated for any $\tau$.
Finally, the lag time was determined as the centroid around the peak
of the CCF, $\tau_{\rm {cent}}$, which was calculated by utilizing
a part of the CCF near the peak with a value larger than
0.8 times the peak value of the CCF.

The uncertainty of the lag time was estimated by Monte Carlo
simulation \citep{maoz+89}.
A pair of artificial light curves was generated for the $V$ and $K$ bands,
and the CCF was calculated to determine the centroid lag,
$\tau_{\rm {cent}}$.
The cross-correlation centroid distribution (CCCD) was then derived
from a large number of realizations of the simulation,
and the uncertainty of the lag time was determined from the CCCD.

Artificial light curves were calculated by the flux randomization (FR)
method \citep{pete+98} combined with a method
based on the structure function (SF),
which simulates the flux variability between the observed data points
\citep{suga+06}.
According to Gaussian distributions with standard deviations
given by the flux-measurement errors for respective data points,
the FR method randomly modifies the observed fluxes in each realization.
The basic concept of simulating the flux uncertainty at interpolated data
points caused by the variability in using SF is described
in the following equation. Here, we define SF as 
\begin{equation}
SF(\tau)=\left(\frac{1}{N(\tau)}\right)\sum_{i<j}[f(t_{i})-f(t_{j})]^{2}-2\sigma^{2}
\end{equation}
where $f(t)$ is the flux at epoch $t$,
$\sigma$ is the average of the observed flux errors of the pair,
and $N(\tau)$ is the number of data pairs;
the sum is given by the overall pairs for which $t_{i}-t_{j}=\tau $.
According to this definition,  $(SF(\tau))^{1/2}$ can be interpreted as
the standard deviation of flux variation between two observational epochs
with a time interval of $\tau$.
Then, an artificial flux for an arbitrary epoch apart from actual
observational epochs can be stochastically realized according to
the Gaussian distribution,
in which the standard deviation is given by $(SF(\tau))^{1/2}$.
A detailed procedure for calculating the artificial light curve
has been reported by \citet{suga+06}.

\begin{figure}
\epsscale{0.6}
\plotone{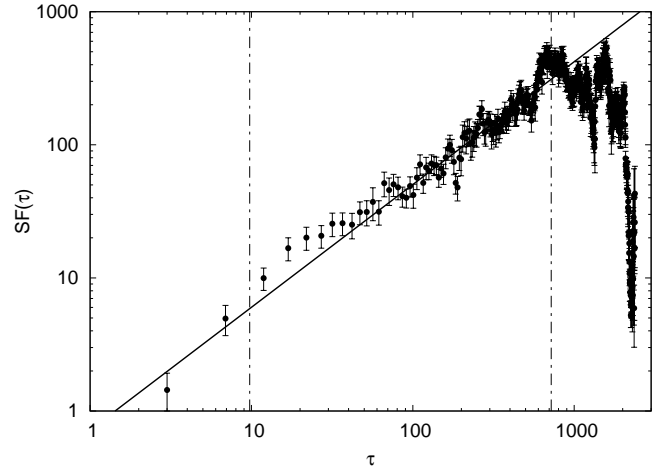}
\caption[The Structure Function]{
Structure function of the $V$-band flux variation of NGC 4151.
The filled circle shows the data point for each bin of the time
interval $\tau$ calculated from the observed light curve;
the solid line shows the regression line in a power-law form
in the range of the time interval
in which the structure function is considered to be significant
(between the two dot-dashed lines).
}\label{fig:sf_ngc4151}
\end{figure}

In Figure \ref{fig:sf_ngc4151}, the $V$-band SF for NGC 4151 is
presented as an example of the SF.
The power-law form of $SF(\tau)=\alpha \tau^{\beta}$ 
was fitted in the range of the time interval between
two times the median of the observational intervals
and $0.3$ times the entire observation span,
in which SF is considered to be significant \citep{coll+01}.
The regression line of the power-law fitting for NGC 4151
is also presented in Figure \ref{fig:sf_ngc4151},
and the parameters of the power-law fitting of the
SFs for all the 17 targets are listed in Table \ref{tab:ana:lc:sf}.
Because the $K$-band light curves were smooth
and appeared to be well sampled,
we applied the SF method described above only for the $V$-band data
\citep{kosh+09}.

We conducted 1000 realizations of the Monte Carlo
simulation for each target to obtain the CCCD.
Realizations were omitted if the significance of
the correlation coefficient at the CCF peak was less than $0.90$
or if the CCF showed a plateau or multiple peaks.
Finally, the lag time of the flux variation of the dust-torus emission
in the $K$ band after that of the emission from the accretion disk,
$\Delta t_K$, was estimated by a median of the CCCD,
and its uncertainty was estimated by 34.1 percentiles
in the upper and lower sides of the median of the CCCD.

\begin{figure}[t]
\epsscale{1.1}
\plotone{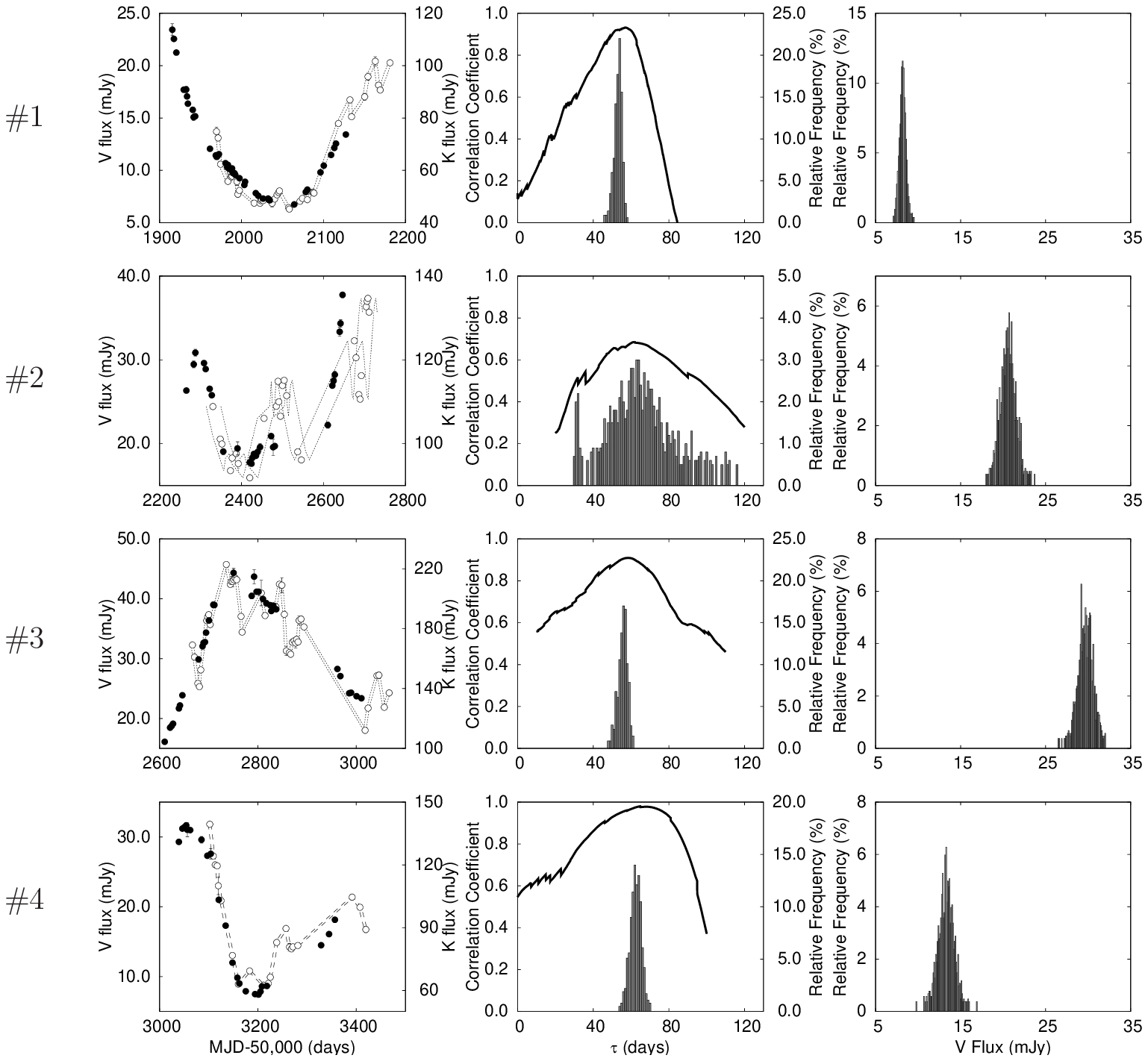}
\caption[The CCF and the CCCD]{
Results of the cross-correlation function (CCF) analysis of NGC 4151
at different epochs (\#1--\#4).
Left panels show the $K$-band (filled circles)
and $V$-band (open circles) light curves.
The latter have shifted according to the lag time between them,
$\Delta t_K$, as derived from the CCF analysis.
Dotted lines show the $V$-band light curves shifted by
$\Delta t_K$ plus and minus its error.
Middle panels show the CCF of the observed light curves (solid line)
and the cross-correlation centroid distribution (CCCD) derived from
Monte Carlo simulation (histogram) for respective epochs.
Right panels show the distribution of
the average $V$-band flux $\bar{f}_V$
derived from the Monte Carlo simulation.
}\label{fig:ccf1}
\end{figure}

\begin{figure}[t]
\epsscale{1.1}
\plotone{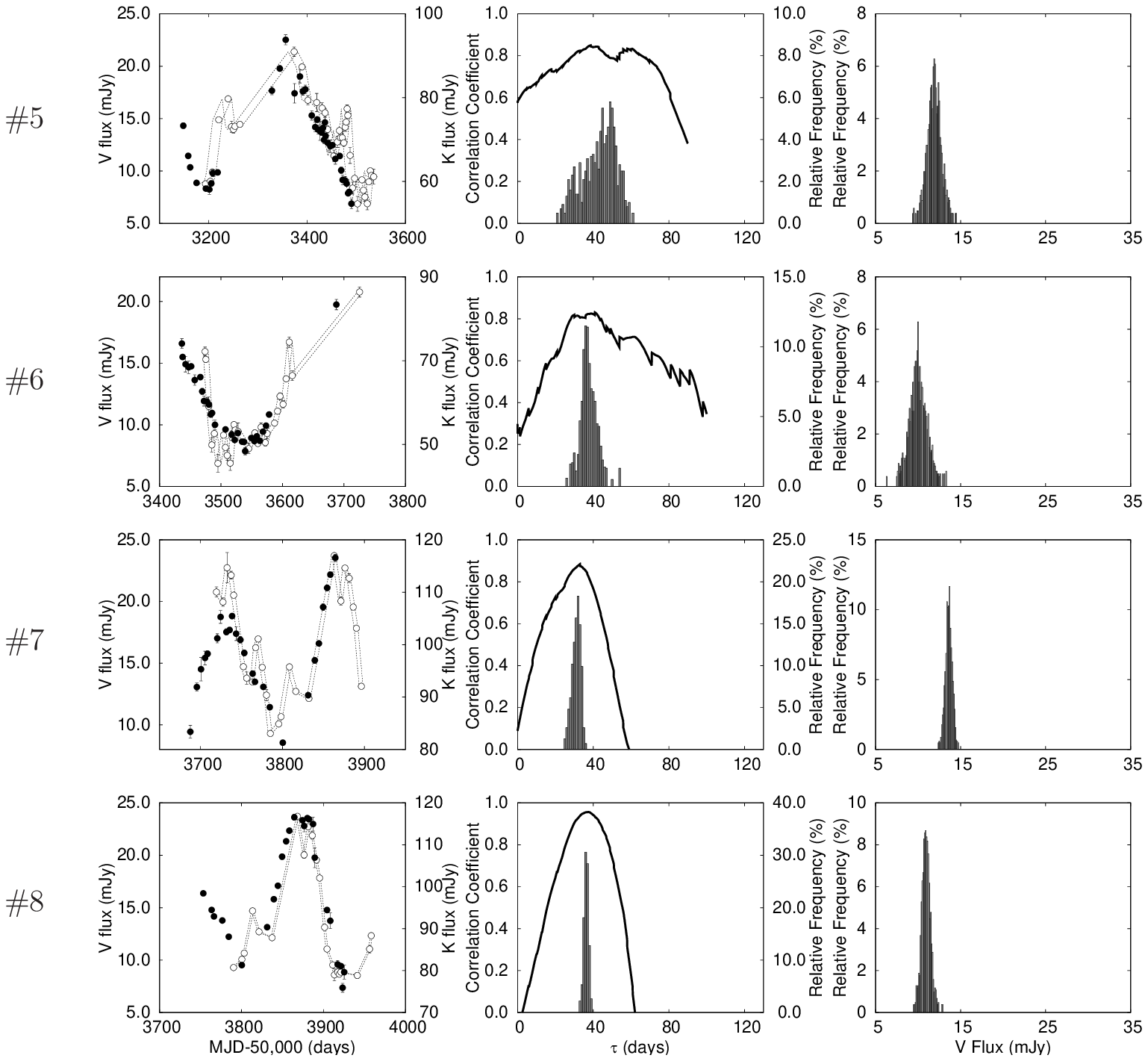}
\caption[The CCF and the CCCD]{
Results of the CCF analysis of NGC 4151 at different epochs
(\#5--\#8).
Symbols and others are the same as Figure \ref{fig:ccf1}.
}\label{fig:ccf1_2}
\end{figure}

As shown in Figures \ref{fig:lc1}--\ref{fig:lc3},
multiple features in the light curves such as local peaks and
valleys were apparent in several AGNs;
we estimated their lag times at different epochs.
We separated the entire monitoring period into several sections,
each of which contained a small number of features of flux variation
and more than 30 data pairs for sufficient calculation of
significant CCFs.
In Figures \ref{fig:ccf1} and \ref{fig:ccf1_2},
CCFs and CCCDs for the individual light curve sections
are presented for NGC 4151 as examples.
As a result, 49 data of $\Delta t_K$ were
obtained for the 17 target AGNs.
The lag times and their errors for all measurements,
in addition to the weighted averages of the lag times
at different epochs for the individual targets,
are listed in Table \ref{tab:dis:dtvar:tab2}.

\begin{figure}
\epsscale{0.7}
\plotone{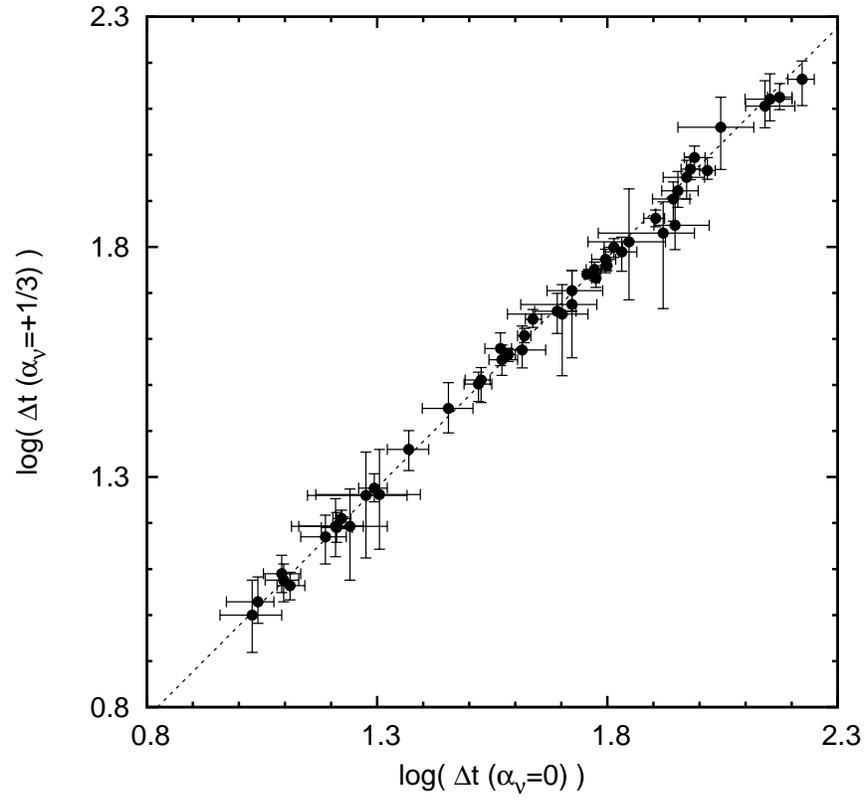}
\caption[comparison between lags]{
Comparison of the dust lags $\Delta t_K$
obtained by the CCF analysis
assuming $\alpha_{\nu}=0$ and $+1/3$ for subtraction of
the accretion-disk component in the $K$-band flux.
The dashed line represents the best-fit regression line.
}
\label{fig:res:lag_lag}
\end{figure}

To compare the lag times $\Delta t_K$,
assuming $\alpha_{\nu}=0$ and $+1/3$ for subtracting
the accretion-disk component in the $K$-band flux,
we plotted all 49 measurements of $\Delta t_K$
in Figure \ref{fig:res:lag_lag}.
The lag times $\Delta t_K$ in which $\alpha_{\nu}=0$ was assumed
and those in which $\alpha_{\nu}=+1/3$ was assumed strongly correlated,
and the best fit linear regression for Figure \ref{fig:res:lag_lag} was
$\log \Delta t_K(\alpha_{\nu}=+1/3)-\log \Delta t_K(\alpha_{\nu}=0)=-0.023\pm 0.005$.
Therefore, the maximum systematic error for $\Delta t_K$ 
caused by the uncertainty of the spectral shape of
the flux variation of accretion-disk emission
in optical and near-infrared spectra
for the subtraction of the accretion-disk component in the $K$-band fluxes
can be estimated as approximately 5\%.

\subsection{Measuring the Dust Lag by JAVELIN}

Next, we applied an alternative method for
measuring lag times developed by \citet{zu+11},
the JAVELIN software (formerly known as SPEAR),
which is employed in recent reverberation studies
\citep{grie+12a,grie+12b,2012ApJ...757...53D,2013ApJ...764...47G,2013MNRAS.431L.112Z,2013MNRAS.434.2664Z,2013A&A...559A..10S}.
Its formalism was originally developed by \citet{rk94} 
based on the studies of \citet{pres92} and \citet{rp92},
and has the advantage that
the uncertainties in the interpolation of the light curve data
and thus the statistical confidence limits on the lag time including them
can be self-consistently estimated
under a statistical model of variability.

JAVELIN adopts a damped random walk (DRW) model
for flux variation of the UV-optical continuum emission,
which has been demonstrated to be a good statistical model for AGN variability
\citep[e.g.,][]{2009ApJ...698..895K,2010ApJ...708..927K,2010ApJ...721.1014M,2012ApJ...753..106M,2013ApJ...765..106Z},
and also assumes a top hat transfer function
in the transfer equation of reverberation mapping.
Then, it fits the continuum and response light curves
using the Markov chain Monte Carlo (MCMC) method
to obtain the likelihood distribution for the lag time as well as
those for the two DRW model parameters for the continuum variability,
the width of the transfer function, and the scaling coefficient
that determines the response for a given change in the continuum.

The parameter values of the DRW model and their uncertainties are supplied
to the lag-time fitting process as prior distributions,
thus they are usually obtained
from the continuum light curve before the lag-time fitting.
In this study, we obtain lag times for separated monitoring sections,
and each of them has less observational data points than those
for the entire monitoring period.
Therefore, 
we first obtained the best-fit values of the DRW model parameters
from the entire $V$-band light curve of each target,
then we supplied those best-fit values with very small uncertainties
to the lag-time fitting process as prior distributions,
thus the DRW model parameters were effectively fixed to those values
during the lag-time fitting process.
This is a similar way to the SF parameters and the CCCD analysis
as described in the previous subsection.

The resultant lag times $\Delta t_K$ obtained by JAVELIN
are also listed in Table \ref{tab:dis:dtvar:tab2}.
In most cases, a single and unambiguous peak of the likelihood
distribution for $\Delta t_K$ could be found.
However, the likelihood distribution was sometimes not stable
depending on the MCMC control parameters.
Additionally, ambiguous multiple peaks sometimes appeared
in the likelihood distribution, some of which
might be regarded as a single but broad peak in the CCCD.
As \citet{zu+11} reported on an aliasing problem of JAVELIN,
it tends to map the $K$-band light-curve data
into the sampling gaps of the $V$-band light curve
to decrease the data overlap,
then poor light curve sampling would lead to
those problems in the lag-time analysis.

\begin{figure}
\epsscale{0.7}
\plotone{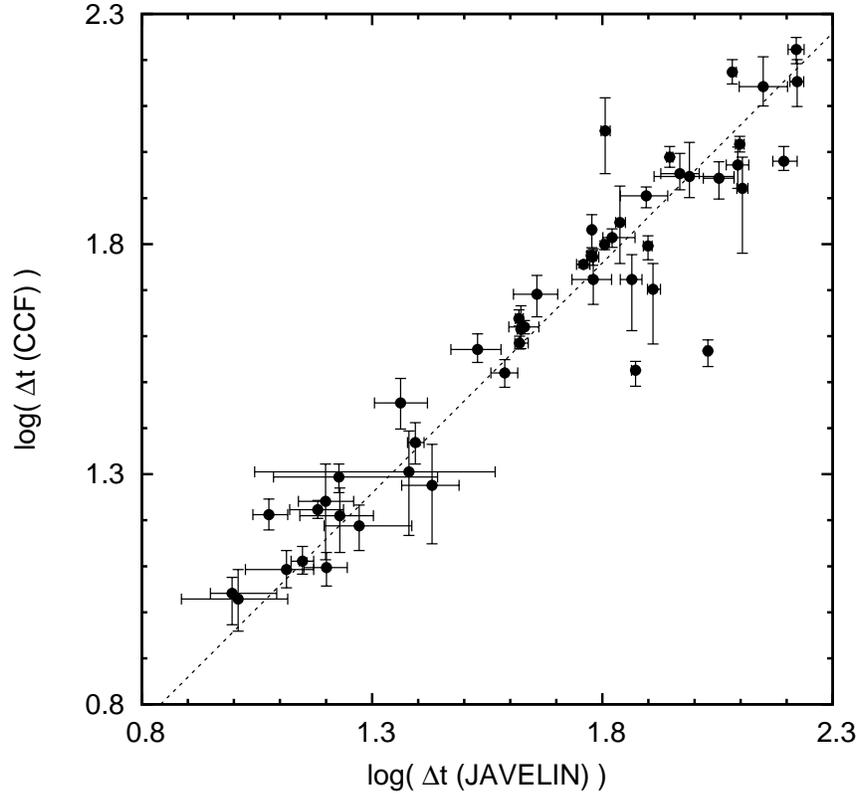}
\caption[comparison between lags]{
Comparison of the dust lags $\Delta t_K$
obtained by using the CCF analysis and JAVELIN
(assuming $\alpha_{\nu}=0$ for the subtraction of
the accretion-disk component in the $K$-band flux).
The dashed line represents the best-fit regression line.
}
\label{fig:CCFJAV}
\end{figure}

In Figure \ref{fig:CCFJAV},
we plotted all 49 measurements of $\Delta t_K$
based on the two different methods
and found that the lag time obtained by JAVELIN is generally
in accordance with that obtained by the CCF analysis,
although the scatter of the data points around the best fit regression
is larger than that expected from their estimated errors.
The best fit linear regressions for them are
$\log \Delta t_K({\rm CCF})-\log \Delta t_K({\rm JAVELIN})= -0.041\pm 0.017$
with an additional scatter of $\sigma _{\Delta t}=0.072$ dex in both directions
for the reduced $\chi ^2$ to achieve unity,
and $-0.049\pm 0.016$ with $\sigma _{\Delta t}=0.067$ dex
for $\alpha_\nu=0$ and $+1/3$, respectively.
Therefore, the systematic difference for $\Delta t_K$
between the two different lag analysis methods
can be estimated as approximately 10\%.
In addition, as will be presented in the next section,
the scatter around the dust lag-luminosity correlation
for $\Delta t_K({\rm JAVELIN})$ is
as much as that for $\Delta t_K({\rm CCF})$.
By these considerations,
we use the $\Delta t_K$ data obtained by using JAVELIN
as well as those obtained by using the CCF analysis
in the subsequent discussions,
although the data plots are represented by
$\Delta t_K(\rm CCF;\ \alpha_{\nu}=0)$ for clarity.

\section{Lag-Luminosity Correlation for the Dust Torus}\label{sec:laglum}

We present 49 measurements of dust reverberation
for the 17 type-1 AGNs, the data for which were analyzed
carefully and uniformly to obtain precise estimates of lag times.
This sample is the largest homogeneous collection
of dust reverberation and observations for the size of the inner dust torus.
By using these results, we examined the correlation
between the dust lag and optical luminosity,
which has been reported in previous studies \citep{suga+06,kish+11}.
In the subsequent sections, the lag times were corrected
for time dilation according to the object redshift,
the maximum of which was approximately 5\%.

\subsection{Estimation of the Optical Luminosities of an AGN}

We first calculated an ``average'' $V$-band flux, $\bar{f}_V$,
for each corresponding period in which the $\Delta t_K$
was obtained, following \citet{kosh+09}.
Rather than obtaining a simple average of the observed fluxes,
which could be biased by unequal intervals of observation,
an average flux from the interpolated $V$-band light curve
with equal intervals generated in the process of the CCF analysis
was calculated as a measure of $\bar{f}_V$.
We obtained the distribution of the average $V$-band flux
from all realizations of the Monte Carlo simulation,
and $\bar{f}_V$ and its error were estimated by 
a centroid and $34.1$ percentiles in both sides of the distribution.
In Figures \ref{fig:ccf1} and \ref{fig:ccf1_2},
the distributions of the average $V$-band flux
for respective epochs are presented for NGC 4151 as an example.

The absolute magnitude in the $V$ band, $M_V$, was then calculated
from $\bar{f}_{V}$ and the recession velocity of the target,
assuming the cosmological parameters.
The recession velocity was obtained from the local flow corrected
velocity of the Virgo infall $+$ Great Attractor $+$ Shapley supercluster
displayed by the NED,
which was originally based on \citet[][]{moul+00}.
The $V$-band absolute magnitude $M_V$ and their errors of the targets
for all epochs of lag-time measurements,
in addition to the weighted averaged $M_V$ 
for the individual targets, are listed
in Table \ref{tab:dis:dtvar:tab2}.
The error of $M_V$ was estimated from a root sum square of
the errors of $\bar{f}_V$, host-galaxy flux, 
narrow emission-line flux, and flux calibration
caused by uncertainty of the reference star magnitudes.

\subsection{Correlation between the Dust Lag and Optical Luminosity}\label{res:corrmvdt}

\begin{figure}
\epsscale{0.7}
\plotone{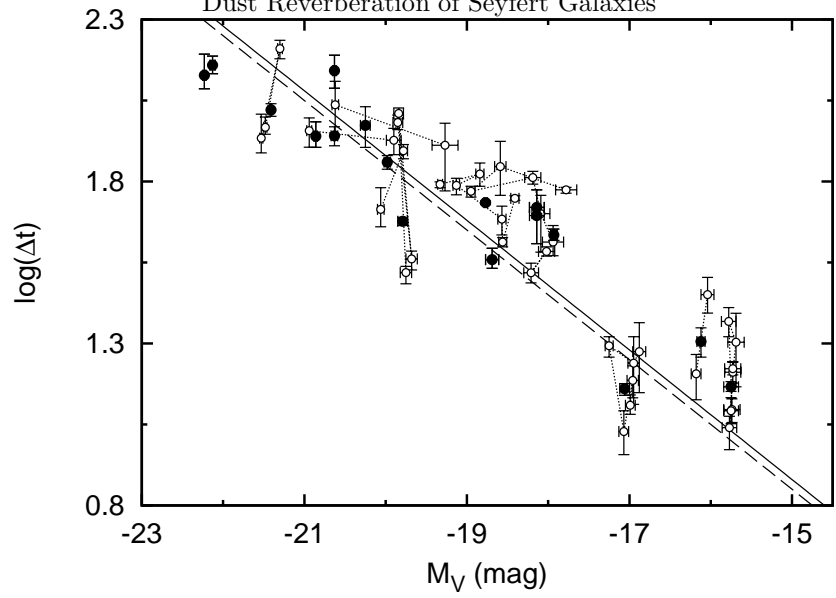}
\caption[The lag-luminosity correlation]{
Lag times between the $V$-band and $K$-band light curves
plotted against the $V$-band absolute magnitude.
Open circles represent 49 measurements of the lag times
(obtained by the CCF analysis
assuming $\alpha_{\nu}=0$ for the subtraction of
the accretion-disk component in the $K$-band flux)
and $V$-band absolute magnitude, and the data points
are connected with dotted lines for the same target.
Filled circles represent the weighted averaged data
for 17 individual target active galactic nuclei (AGNs).
Solid and dashed lines represent the best-fit regression lines
for the data obtained by the CCF analysis
assuming $\alpha _{\nu}=0$ and $\alpha _{\nu}=1/3$ respectively.
The lag times were corrected for the time dilation according to the object redshift.
}
\label{fig:dis:mvdt:mvdt}
\end{figure}

In Figure \ref{fig:dis:mvdt:mvdt},
the dust lags $\Delta t_K$ of the target AGNs were plotted
against their $V$-band absolute magnitudes $M_V$.
A clear correlation between $\Delta t_K$ and $M_V$
over an absolute magnitude range of $M_V\approx -16$ to $-22$
was detected for the 49 measurements of dust lags
and the weighted averaged lags for the 17 individual target AGNs,
represented in the figure by open and filled circles, respectively.
This correlation has been presented in previous studies
for a smaller sample \citep{okny+01,mine+04,suga+06}.

To examine the correlation between the dust lag
and optical luminosity,
a regression analysis was applied with a linear equation
in the form of
\begin{equation}
\log \Delta t_K=a+b M_{V}\, . \label{eq:result1}
\end{equation}
This regression analysis was applied to the data
assuming $\alpha_{\nu}=0$ and $+1/3$
for both $\Delta t_K({\rm JAVELIN})$ and $\Delta t_K({\rm CCF})$.
We assumed the slope of the regression line as $b=-0.2$,
following \citet{mine+04},
because the slope is expected from the simple model of
the dust sublimation radius, as presented in Equation (\ref{eq:intro:1}).
In addition, \citet{kish+11} reported that such correlation
is consistent with the slope for the data combined with
the dust reverberation and near-infrared interferometry,
including AGNs with higher luminosity.
The regression was calculated by
the generalized multivariate least-squares method
presented by \citet{jeff80},
weighted by the errors of $\Delta t_K$ and $M_V$.
A detailed discussion on the slope of the correlation
will be available in a forthcoming paper
based on the systematic measurements of dust reverberation
with the data for AGNs with higher luminosity.

We first fit the model
to the data of the $49$ measurements of dust lags
and to those of the weighted averaged lags for the 17 targets;
the results are listed in Table \ref{tab:mvdtfit}.
The derived parameters agreed well;
however, the reduced $\chi ^2$ of
the fitting was significantly larger than unity.
That is, the fitting residuals were much larger than
those expected from the measurement errors of
$\Delta t_K$ and $M_V$.

Peculiar velocity was a source of error
for our estimation of $M_V$,
whereby the effect was larger for closer objects.
Since the values presented for the peculiar velocity
dispersion, $\sigma_{\rm{pec}}^2$,
differ slightly \citep[e.g.,][]{watk97,gibb+01,radb+04,mast+06},
we assumed $\sigma_{\rm{pec}}=200$ km s$^{-1}$
and $300$ km s$^{-1}$ to estimate the uncertainty
of $M_{V}$ caused by peculiar velocity.
Then, incorporating this $M_{V}$ error
by root-sum-square,
we fit the model to the data of the weighted averaged lags
for the 17 targets.
These results are also listed in Table \ref{tab:mvdtfit}.
The reduced $\chi ^2$ of the fitting decreased considerably
when the peculiar velocity dispersion was incorporated;
however, it remained much larger than unity.
Therefore, significant intrinsic scatter
was expected to reside in the dust lag-luminosity
correlation.

We then fitted the model to the data of the weighted averaged lags
for the 17 targets, adding an error $\sigma_{\rm{add}}$ to the error
of $\log \Delta t_K$ data by root-sum-square
for the reduced $\chi^2$ to achieve unity.
The fitting was also applied to the data incorporating
the peculiar velocity dispersion; the fitting results
as well as $\sigma_{\rm{add}}$ are listed in Table \ref{tab:mvdtfit}.
The derived parameters $a$ and $\sigma_{\rm{add}}$ agreed well among them,
and we adopted
$a=-2.12\pm 0.04$ with $\sigma_{\rm{add}}=0.14$ dex ($\alpha_{\nu}=0$ assumed) and
$a=-2.15\pm 0.04$ with $\sigma_{\rm{add}}=0.14$ dex ($\alpha_{\nu}=+1/3$ assumed)
for $\Delta t_K({\rm CCF})$,
and
$a=-2.08\pm 0.04$ with $\sigma_{\rm{add}}=0.16$ dex ($\alpha_{\nu}=0$ assumed) and
$a=-2.10\pm 0.04$ with $\sigma_{\rm{add}}=0.16$ dex ($\alpha_{\nu}=+1/3$ assumed)
for $\Delta t_K({\rm JAVELIN})$,
from the fittings incorporating the peculiar velocity dispersion.
The best-fit regression lines are also presented in Figure \ref{fig:dis:mvdt:mvdt}.
The systematic difference between two linear regressions
between the data assuming $\alpha_{\nu}=0$ and $+1/3$ was $0.02-0.03$ dex,
and that between the CCF analysis and JAVELIN was $0.04-0.05$ dex. 
We adopted $\alpha_{\nu}= +0.10$ \citep{tomi05} tentatively 
for the spectral shape of the flux variation of accretion-disk
emission in optical and near-infrared spectra for
the subtraction of the accretion disk, and we estimated
$a=-2.13\pm 0.04$ for $\Delta t_K({\rm CCF})$
and $a=-2.09\pm 0.04$ for $\Delta t_K({\rm JAVELIN})$.
Then, we adopt $a=-2.11\pm 0.04$ for the parameter
of the dust lag-luminosity correlation in Equation (4).
The parameter of $a=-2.11\pm 0.04$ was consistent with
that presented in a previous study \citep[$a=-2.15$;][]{mine+04}.
In addition,
regarding the scatter $\sigma_{\Delta t}$
between $\Delta t_K({\rm CCF})$ and $\Delta t_K({\rm JAVELIN})$
as the systematic scatter associated with the lag-analysis methods,
$\sigma_{\rm{int}}=\sqrt{\sigma_{\rm add}^2-\sigma_{\Delta t}^2}\approx 0.13$ dex
can be interpreted
as intrinsic scatter in the dust lag-luminosity correlation.

\subsection{Possible Origin of Intrinsic Scatter in the Dust Lag-Luminosity Correlation}\label{res:corrmvdtscatter}

Although the central engine of the type-1 AGN is not considered
to be as heavily obscured by the dust torus as that of the type-2 AGN,
a small but different amount of extinction could exist
in type-1 AGNs to create a possible source of scatter
in the dust lag-luminosity correlation.
\citet{cack+07} estimated the intrinsic color excess of 14 AGNs
by the flux variation gradient (FVG) method \citep{wink+92}
and the Balmer decrement method.
The estimates obtained from both methods were consistent; therefore,
the difference between the estimates by the two methods
could be regarded as the sum of the intrinsic uncertainties of
the methods themselves.
We estimated the standard deviation of
the uncertainty of the FVG method as $0.041$ mag.
In fact, of the 14 target AGNs of \citet{cack+07},
11 AGNs are in common with our targets.
For these targets,
the standard deviation of color excess obtained by the FVG method
was estimated as $\sigma _{E(B-V),\rm{FVG}}=0.076$ mag,
and the intrinsic standard deviation of color excess
was estimated as $\sigma _{E(B-V)}=0.064$ mag.
Assuming $A_V/E(B-V)=3.1$, this value corresponds
to the scatter of approximately $0.2$ mag
in $M_V$ or $0.04$ dex in $\log \Delta t_K$,
which is insufficient for explaining the scatter of
$\sigma_{\rm{int}}\approx 0.13$ dex in the correlation.

The difference in the spectral shape of the flux variation
of accretion-disk emission between targets,
which would be caused by a different amount of 
intrinsic extinction or by other reasons, would contribute
the scatter in the dust lag-luminosity correlation.
For example, the intrinsic standard deviation
of $\sigma _{\alpha _{\nu}}= 0.1$,
which could be caused by the intrinsic standard deviation
of color excess of $\sigma _{E(B-V)}=0.054$ mag,
would originate the scatter of approximately $0.01$ dex in $\log \Delta t_K$.
This result is also insufficient for explaining the scatter of
$\sigma_{\rm{int}}\approx 0.13$ dex in the correlation.

We then focused on the changes in dust torus size with the flux variation
as an additional possible source of scatter
in the dust lag-luminosity correlation.
\citet{kosh+09} found that although the dust lag changed with
the flux variation in the $V$ band, the dust lags
at different epochs did not follow the relation of
$\Delta t \propto L_V^{0.5}$ for NGC 4151.
They suggested that the destruction and formation of
the dusty clouds composing the dust torus would not
respond instantaneously with the variation of incident flux.
In addition, \citet{pott+10} and \citet{kish+11} did not detect
the size change of the inner dust torus during the flux variation
in the multi-epoch near-infrared interferometric data for the same object,
and at last \citet{2013ApJ...775L..36K} found
its size change from the longer term monitoring observation by
the near-infrared interferometry, which
did not follow the relation of
$\Delta t \propto L_V^{0.5}$ as well.
Indeed, as shown in Figure \ref{fig:dis:mvdt:mvdt},
the multi-epoch data of $\Delta t_K$ and $M_V$
for a target AGN appeared to scatter around
the global dust lag-luminosity correlation.
This behavior of the dust torus size would
broaden the correlation 
unless the dust lag and luminosity for an object
were time-averaged over a long span.
The scatter was then estimated
from the multi-epoch data for the 11 target AGNs.
The median of the standard deviations of distance
from the $\Delta t_K \propto L_V^{0.5}$ correlation
for respective targets was 0.11 dex, 
which is significant but still has some room to account for
the $\sigma_{\rm{int}}\approx 0.13$ dex scatter.

Even after these possible contributions to the scatter were subtracted,
an intrinsic scatter of $\sigma \lesssim 0.1$ dex in $\log \Delta t_K$
was expected to remain in the dust lag-luminosity correlation
and would be produced by diversities of characteristics in type-1 AGNs
such as the viewing angle of the accretion disk and dust torus
or by the Eddington ratio of the mass accretion rate.
The dependency of dust lag
on these parameters is discussed in Section \ref{sec:dis}.

\section{Discussion}\label{sec:dis}

\subsection{Change in Dust Lag with Viewing Angle and Mass Accretion Rate}

The dust reverberation models predicted that
the transfer function,
a response of flux variation of the dust emission
on an impulse of the UV-optical continuum emission,
would change with the viewing angle of the dust torus
\citep{barv92,kawa+11}.
In particular, \citet{kawa+11} reported that
the centroid of the transfer function,
which should be interpreted as the lag between flux variations,
increases with the viewing angle according to
the reverberation model with a clumpy torus,
because the self-occultation and waning effects
selectively decrease the emission from dusty clouds
with a shorter lag at the side in closest proximity to the observer.

\begin{figure}
\epsscale{0.7}
\plotone{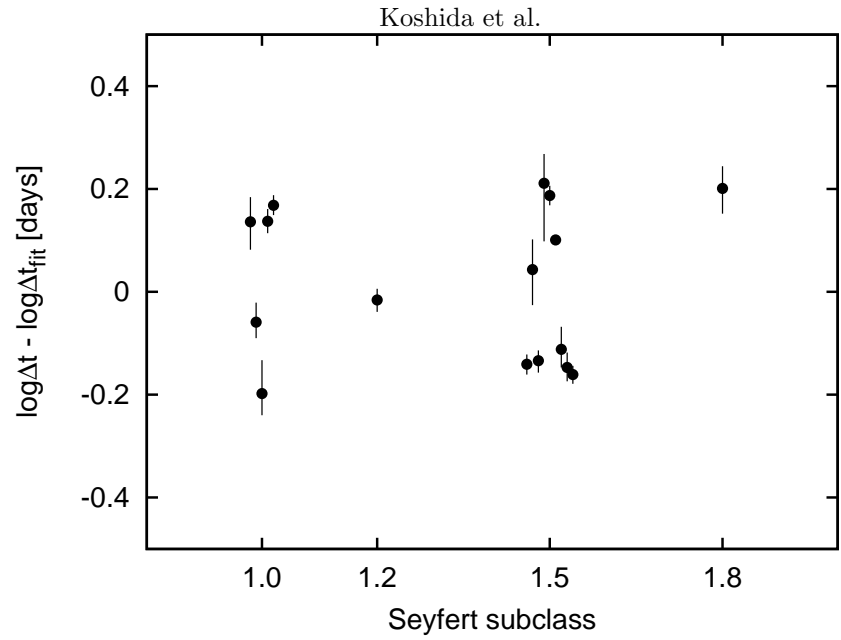}
\caption[]{
Residual of the dust lag from
the best-fit regression line plotted against 
the subclass of the Seyfert type
(obtained by the CCF analysis
assuming $\alpha_{\nu}=0$ for the subtraction of
the accretion-disk component in the $K$-band flux).
}
\label{fig_sytypdt}
\end{figure}

We then examined a possible systematic change
in dust lag with the viewing angle.
In Figure \ref{fig_sytypdt}, we plotted the residuals
of the dust lag from the best-fit regression line
(for the data assuming $\alpha_\nu =0$)
against the subclass of the Seyfert type
as an indicator of viewing angle.
The subclasses of intermediate Seyfert galaxies
(S1.0, S1.2, S1.5, and S1.8) for the 14 target AGNs
were obtained from \citet{vero+10}.
Among them, Mrk 335 and NGC 4051
were classified as narrow-line Seyfert 1 (S1n)
by \citet{vero+10},
and a subclass of S1.0 was assigned
following \citet{oste+93}. 
The error of the residual was estimated from
those of the lag time and $M_V$.

As shown in Figure \ref{fig_sytypdt},
we did not detect systematic changes of dust lag
with the subclass of the Seyfert type,
although \citet{kawa+11} estimated that
the dust lag systematically shifted
approximately $0.4$ dex with a change of
$0^{\circ }-45^{\circ }$ in the viewing angle.
However, a possible correlation between the dust lag
and viewing angle could not be ruled out,
because a certain amount of scatter in the relationship
between the Seyfert type and viewing angle
was expected from a realistic clumpy dust torus model,
which consists of dusty clouds stochastically
distributed following the probability distribution
as a function of the angle from the equatorial plane
\citep{elit12}.
A better estimate of the viewing angle is
required for further investigation.

Next, we focused on the Eddington ratio,
which is the mass accretion rate
relative to the mass of the central black hole.
When the mass accretion rate exceeds the Eddington rate,
a slim disk that is optically and geometrically
thick appears \citep{abra+88}.
Its disk illumination toward the dust torus
located on the equatorial plane would be suppressed
by disk self-occultation \citep{kawa+11,kaoh11};
therefore, we can expect that the inner radius of the dust torus
decreases as the mass accretion rate increases
according to the stronger anisotropy of disk illumination.
However, \citet{kawa+11} reported that
according to their reverberation model,
the change in dust lag would not be as significant
as the mass accretion rate.

\begin{figure}
\epsscale{0.7}
\plotone{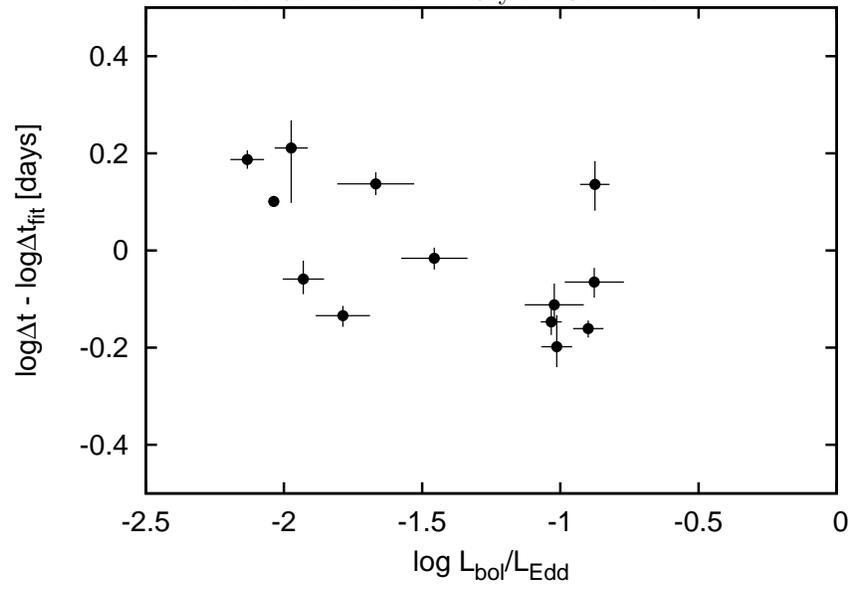}
\caption[]{
Residual of the dust lag from
the best-fit regression line plotted against the Eddington ratio
(obtained by the CCF analysis
assuming $\alpha_{\nu}=0$ for the subtraction of
the accretion-disk component in the $K$-band flux).
}
\label{fig_erdt}
\end{figure}

In Figure \ref{fig_erdt}, we plotted the residuals
of the dust lag from the best-fit regression line
(for the data assuming $\alpha_\nu =0$)
against the Eddington ratio, $L_{\rm bol}/L_{\rm Edd}$,
for the 13 target AGNs with black-hole masses
measured by BLR reverberation mapping
\citep[][]{pete+04,bent+06a,bent+07,denn+10,grie+12b}.
NGC 4593 was omitted because of its large error in
black-hole mass.
The bolometric luminosity was estimated from
the luminosity of the optical continuum emission as
$L_{\rm bol}=9\times \lambda L_{\lambda} (5100\mathrm{\AA})$
\citep{kasp+00}, where $L_{\lambda} (5100\mathrm{\AA})$ 
was calculated from $L_V$ by assuming
a power-law spectrum of $f_{\nu}\propto \nu^0$.
The error of the Eddington ratio was estimated from
those of $M_V$ and black-hole mass.

As shown in Figure \ref{fig_erdt},
we did not detect systematic changes in dust lag
with the Eddington ratio.
In fact, the Eddington ratios of all 13 targets
were smaller than $\sim 0.1$
and were believed to be too small for slim disk formation.
Because a standard accretion disk with a small mass accretion rate
would not change the illumination anisotropy
as much as the mass accretion rate \citep{kawa+11},
the absence of a systematic trend of dust lag with the Eddington ratio
in Figure \ref{fig_erdt} is consistent with the model.
The observation of an AGN with a super-Eddington mass accretion rate
is required to investigate the effect of
anisotropic illumination of the slim disk.

\subsection{Structure of the BLR and Inner Dust Torus}
Reverberation observations for BELs and dust thermal emission
provide an important tool for investigating the structure
of the BLR and innermost dust torus.
\citet{suga+06} compared the results of the dust reverberation
with those for BELs and demonstrated that the innermost dust torus
was located just outside of the BLR.
Infrared interferometry is also an important tool
for investigating the dust torus.
\citet{kish+11} reported that
the radius of the innermost dust torus
obtained by near-infrared 
interferometry was consistent with
the dust reverberation radius;
however, the former tended to be roughly equal to or
slightly larger than the latter.
On the basis of our new results of the largest 
homogeneous sample
of dust reverberation, we discuss in this section
the structure of the BLR and the innermost dust torus.

In Figure \ref{fig:dis:mvdt:lrv},
we plotted the radii of the innermost dust torus
obtained by reverberation and near-infrared interferometry,
as well as the reverberation radius of the BLR,
against the optical $V$-band luminosity.
The dust reverberation radii and optical luminosities
were obtained from our results,
and those of the near-infrared
interferometry
were obtained from the data compiled by \citet[][]{kish+11} and
\citet[][]{weigelt+12}.
The lag-luminosity correlation of BLR has been investigated
by many authors. Among the data presented in such studies,
we used that obtained by the reverberation observation
of Balmer emission lines (mostly H$\beta $)
compiled by \citet{bent+09a},
because they presented an accurate estimation
of the optical luminosity by using the $HST$ images
to carefully subtract the host-galaxy flux.

In Figure \ref{fig:dis:mvdt:lrxo},
the radii of the innermost dust torus and BLR (H$\beta $)
were plotted against the hard X-ray (14--195 keV) luminosity,
and in Figure \ref{fig:dis:mvdt:lrxo_2},
they were plotted against the luminosity
of [\ion{O}{4}]$\lambda 25.89$ $\mu$m emission line.
Although the hard X-ray emission
and [\ion{O}{4}] emission line
are not directly related to 
the ionization state of the BLR clouds
or dust temperature,
their luminosities are expected to correlate with
that of the accretion disk emission,
and they would be far less obscured by the dust torus.
Therefore, they could serve as luminosity indicators
unbiased to dust obscuration
\citep[e.g.,][]{mele+08,diam+09,rigb+09}
and would be useful for estimating the radii of the BLR
and innermost dust torus for type-2 AGNs and
ultraluminous infrared galaxies.
We enlarged the data for reverberation radii of the BLR
by adding the results of the recent reverberation observations
for H$\beta $ emission lines
\citep[][]{bent+06b,bent+07,bent+09a,bent+09b,denn+10,bart+11a,bart+11b,grie+12b}.
The hard X-ray luminosity was obtained from
the Swift BAT 58-Month Hard X-ray Survey
(Baumgartner et al. submitted)
\footnote{
http://heasarc.gsfc.nasa.gov/docs/swift/results/bs58mon/
},
and that of the [\ion{O}{4}] emission line
was taken from previous research \citep{deo+07,mele+08,diam+09,gall+10,gree+10,liuw+10,tomm+10}.

As shown in Figures \ref{fig:dis:mvdt:lrv}--\ref{fig:dis:mvdt:lrxo_2},
the radii of the BLR (H$\beta $) and innermost dust torus
showed significant correlation with these luminosities,
and the reverberation radius of the BLR was found to be
systematically smaller than that of
the innermost dust torus, as presented by \citet{suga+06}.
In addition, we found that the reverberation radius of
the innermost dust torus appeared to be
systematically smaller than the interferometric radius
of the innermost dust torus,
as presented by \citet{kish+11}.
We then applied the regression analysis for the radius-luminosity
relationship to quantitatively estimate their differences.

For the regression analysis, a linear equation,
\begin{equation}
\log r = \alpha + \beta\log L\, ,
\end{equation}
was adopted and fitted to the data, assuming $\beta=0.5$
in the same manner as that described in Section 4.
We omitted the data for 3C 273, the most luminous target
in Figures \ref{fig:dis:mvdt:lrv}--\ref{fig:dis:mvdt:lrxo_2},
from the fitting with the hard X-ray luminosities,
because it is sometimes classified as a blazer,
and its hard X-ray is apparently more luminous than
the extension of the correlation of the interferometric
radius of the innermost dust torus\footnote{
Because \citet{kish+11} noted that 
3C 273 was in a quiescent state at the observing epoch
of near-infrared interferometry
and that the contribution of the synchrotron emission 
was expected to be small in the optical spectra,
we included 3C 273 for calculating regressions
of the radii with optical luminosity.
}.
Rather than fitting the data, the regression equation for
the correlation of the dust reverberation radius and optical luminosity
was converted from the result given in Section 4,
and that for the correlation of the BLR reverberation radius
and the optical luminosity was obtained from \citet{bent+09a}.
In addition, the regression line for the BLR radius
and the [\ion{O}{4}] luminosity independently 
fitted by \citet{gree+10} are presented for comparison.

The fitted parameters are presented in Table \ref{tab:rlfit}.
We found that the BLR reverberation radius
was smaller than the dust reverberation radius
by $0.6$--$0.7$ dex, or a factor of $4$--$5$,
and that the dust reverberation radius was smaller than
the interferometric radius of the innermost dust torus
by $0.2$--$0.4$ dex, or about a factor of two.

The difference between the reverberation radius
and interferometric radius of the dust torus 
observed in the same band can be understood
by examining the difference between the response-weighted
and flux-weighted radii \citep{kish+11}.
Because the dust temperature in the torus is the
highest at the inner boundary of the dust torus
and becomes lower at larger radii,
the flux-weighted radius would be larger than
the inner boundary radius of the dust torus
caused by the flux contribution
from lower temperature dust at larger radii.
In contrast, the reverberation radius analyzed by
the CCF analysis would be more weighted on a larger amplitude
of flux variation, which is expected to originate
from a more compact emitting region or at smaller radii
in the dust torus.
Therefore, the reverberation radius is expected
to be a better estimate for the inner boundary radius
of the dust torus.

In addition to the factor of $4$--$5$ difference
for the mean reverberation radii of the innermost
dust torus and BLR, a gap between their distributions
is shown in Figure \ref{fig:dis:mvdt:lrv}.
However, on the basis of the following discussions,
we suggest that these data do not necessarily
indicate a gap of matter between the dust torus and the BLR.

The ionized gas clouds in the BLR are considered
to be extended larger than the reverberation radius
of Balmer emission lines.
\citet{clav+91} conducted the reverberation mapping
of various BELs in the UV spectrum for NGC 5548
and found that the lags tended to be shorter for higher
ionization lines and longer for lower ionization lines.
The lag times for the \ion{Si}{4}$+$\ion{O}{4}],
\ion{C}{3}], and \ion{Mg}{2} emission lines
were larger by a factor of two or more
for H$\beta $ in similar epoch observations
\citep{pete+04}.
In addition, \citet{hu+08} and \citet{zhu+09} recently
proposed a component of intermediate velocity width
in conventional broad Balmer emission lines
and suggested that the intermediate line region
is located on the outer part of the BLR
in the vicinity of the dust torus.
Moreover, the ionized gas clouds of the BLR
could be distributed at radii larger than the reverberation radius
because the lag time corresponds to the centroid of the transfer function,
which is considered to be the average radius of
the distribution of the gas clouds
\citep[e.g.,][]{horn99}.

On the contrary, the dust with temperature higher than
that of the black body radiation corresponding to
the $K$-band wavelength
is expected to be in a more inner location than
that of the dust reverberation radius measured in the $K$ band.
\citet{tomi+06} used the multicomponent fitting of the flux variation
to find that the $H$-band flux variation for MCG $+$08$-$11$-$011
was followed by the $K$-band flux variation approximately 6 days later.
Their results indicate that the reverberation radius in the $H$ band 
was approximately 7\% smaller than that in the $K$ band obtained in this study.
The same trend was reported by \citet{tomi05} for additional AGNs.
The $I$-band flux from the higher temperature of dust
was detected by \citet{saka+10},
who also used the multicomponent fitting of the flux variation.
The dust temperature was estimated to be
approximately $1700$--$2000$ K for seven Seyfert galaxies,
which is consistent with the sublimation temperature
of graphite grains \citep{salp77}.

Further, \citet{kosh+09} discovered a rapid variation
of dust lags obtained by the dust reverberation observation 
of NGC 4151. Because this variation was found to be
significantly faster than the dynamical timescale
at the radii of the BLR and inner dust torus,
they concluded that the formation and destruction
of dust grains in the gas clouds changed the radius
of the innermost dust torus.
These results suggest that the gas clouds are
distributed continuously and that the inner boundary
of the dust torus is determined by
the sublimation of the hottest grains in the clouds
\citep{netz93},
which were located between the reverberation radii of
the $K$-band dust emission region and the H$\beta $ BLR.

\citet{mor+09} reported near-infrared excess
emission in the infrared SEDs of AGNs that could not
be explained by the components of emission
from a clumpy dust torus model \citep{nenk+02,nenk+08}
or by the dusty narrow emission-line region (NLR).
\citet{mor+12} introduced a gas cloud
with hot pure-graphite grains to account for 
such excess emission and constructed an SED model consisting
of a hot-dust cloud, a clumpy dust torus,
and a dusty NLR to fit the infrared SED of type-1 AGNs.
Figure \ref{fig:dis:mvdt:lrv} shows the location
of the hot-dust clouds obtained from such fittings
reported by \citet{mor+12}.
Indeed, the hot dust clouds were located
between the two reverberation radii,
which also suggests that
the inner radius of the dust torus
is smaller than the $K$-band dust reverberation radius
but larger than the Balmer-line reverberation radius.

The inner radius of the dust torus
was expected to be smaller than that of
Equation (\ref{eq:intro:1}), assuming
$T_{\rm sub}=1500$ K and $a=0.05\ \mu$m,
by a factor of more than $\sim 3$.
However,
a higher sublimation temperature is probable as described.
In addition,
a larger grain size of $a\approx 0.1\ \mu$m could be adopted
on the basis of the analysis of reddening curves for quasars \citep{gask+04}.
By assuming $T_{\rm dust}=1700$ K for the dust temperature
that was evaluated from the NIR colors of the variable flux component
for Seyfert 1 galaxies \citep{tomi05,tomi+06},
and $a= 0.1\ \mu$m for the grain size,
the sublimation radius predicted by Equation (\ref{eq:intro:1})
can be written as
$\log R_{{\rm sub}}/{\rm pc}=-0.80+0.5\log \ (L_{V}/10^{44}\ \rm{erg\ s}^{-1})$,
where $L_{\rm {UV}}=6L_{V}$ \citep{kish+07},
which was found to be close to
the $K$-band dust reverberation radius.
It should be noted that $L_{\rm {UV}}$ in Equation (\ref{eq:intro:1})
is an effective luminosity of the UV-optical
continuum emission multiplied by the wavelength-dependent
absorption efficiency of dust grains.
Therefore, a detailed calculation of radiation equilibrium
in dusty clouds is necessary
and will appear in a forthcoming paper.

\section{Summary}\label{sec:sum}

We presented the results of a dust reverberation survey
for 17 nearby Seyfert 1 galaxies, which provides
the largest homogeneous data collection for the radius
of the innermost dust torus.
For all targets, long-term monitoring observations
in optical and near-infrared wavelengths showed a delayed response
of the $K$-band light curve after that of the $V$ band.
The minor contribution of the accretion disk
in the $K$-band flux was subtracted to derive
the variation of the dust torus emission,
and the CCF analysis was applied to obtain 49 measurements
of reverberation lags for the innermost dust torus.
The optical luminosity of the continuum emission
from the accretion disk was estimated by
subtracting the host-galaxy and narrow-line flux contributions
in the $V$-band fluxes.
We found that the reverberation lags for the innermost dust torus
strongly correlated with the optical luminosity
in the range of $M_{V}=-16$ to $-22$ mag.
We applied the regression analysis to our new data
to obtain the correlation of
$\log \Delta t\ ({\rm days})\ = -2.11 -0.2M_{V}$
or
$\log c\Delta t\ ({\rm pc})\ = -0.88  +0.5(L_{V}/10^{44} \rm{ergs\ s}^{-1})$,
assuming the slope of $\Delta t \propto L^{0.5}$
to be same as that reported in previous studies.
The intrinsic scatter of the correlation
was estimated to be approximately $0.13$ dex,
and its possible origins were discussed.
The difference in internal extinction
and the delayed response of changes in lag times
to the flux variations could have partly contributed
to the intrinsic scatter;
however, we could not detect systematic changes in
the lag-luminosity correlation
with the subclass of the Seyfert type and the Eddington ratio.

Furthermore, we compared our results with
the radius-luminosity correlations
for the near-infrared interferometry
and BLR reverberation.
In addition to that with the optical luminosity,
we examined such correlations with the hard X-ray (14--195 keV) and
[\ion{O}{4}]$\lambda 25.89$ $\mu$m emission-line luminosities,
which are known as isotropic luminosity indicators
and are applicable to obscured AGNs.
We found that the interferometric radius in the $K$ band was
systematically larger than the reverberation radius in the same band
by approximately 0.3 dex,
which could be interpreted as the difference
between the flux-weighted radius and the response-weighted radius,
as suggested by \citet{kish+11}.
We suggest that the reverberation radius provides a better estimate
for the inner boundary radius of the dust torus.

As expected from the unified scheme of the Seyfert type of AGNs,
the BLR reverberation radius was found to be systematically smaller than 
the dust reverberation radius, as reported by \citet{suga+06},
with a difference of approximately 0.6-0.7 dex.
However, we suggest that
gas clouds with the hottest dust, and those emitting
BELs of lower ionization species 
or intermediate lines, are located between them,
wherein the inner radius of the dust torus is
determined by the sublimation of dust grains.
The higher dust temperature of $T_{\rm dust}\approx 1700$ K
and the larger grain size of $a\approx 0.1\ \mu$m
are preferred for the parameters
of the dust sublimation radius expressed in Equation (\ref{eq:intro:1});
however, detailed calculation of radiation equilibrium
in dusty clouds is desired.

\acknowledgments

We thank the staff at the Haleakala Observatories for their help
with facility maintenance.  This research has been partly supported
by the Grant-in-Aids of Scientific Research
(10041110, 10304014, 11740120, 12640233, 14047206, 14253001,
14540223, 16740106, 22540247, 25287062, and the COE Research (07CE2002)
of the Ministry of Education, Science, Culture and Sports of Japan.



\clearpage

\clearpage

\clearpage

\clearpage

\clearpage

\clearpage

\clearpage

\clearpage

\clearpage

\clearpage

\clearpage

\clearpage

%

\clearpage
\setcounter{figure}{12}
\begin{figure}
\epsscale{0.63}
\plotone{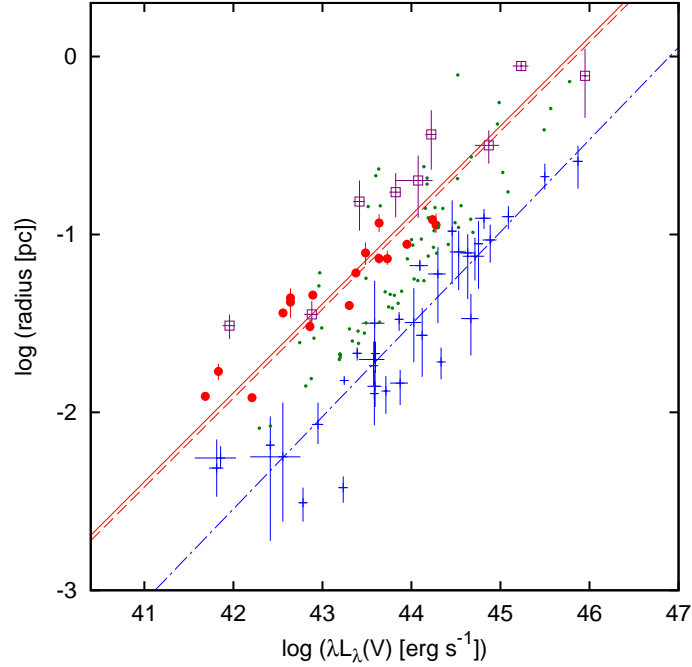}
\caption[Radii of the innermost dust torus and the BLR against optical luminosity]{
Radii of the innermost dust torus and the broad emission-line region (BLR)
plotted against the $V$-band luminosity.
Filled circles (colored red in the online version)
represent the $K$-band reverberation radii of our results
(obtained by the CCF analysis
assuming $\alpha_{\nu}=0$ for the subtraction of
the accretion-disk component in the $K$-band flux);
open squares (colored purple in the online version)
represent the $K$-band interferometric radii
obtained from \citet{kish+11} and \citet{weigelt+12},
and crosses (colored blue in the online version)
represent the reverberation radii
of broad Balmer emission lines obtained from \citet{bent+09a}.
Solid and dashed lines (colored red in the online version)
represent the best-fit regression lines for the $K$-band reverberation radii
for the data obtained by the CCF analysis
assuming $\alpha _{\nu}=0$ and $\alpha _{\nu}=1/3$, respectively,
and the dot-dashed line (colored blue in the online version) 
represents the best-fit regression line
for the Balmer-line reverberation radii reported by \citet{bent+09a}.
Dots (colored green in the online version)
represent the radii of the location of the hot-dust clouds
obtained from the spectral energy distribution (SED)
fitting of type-1 active galactic nuclei (AGNs)
reported by \citet{mor+12}.}
\label{fig:dis:mvdt:lrv}
\end{figure}

%

\clearpage

\setcounter{figure}{13}
\begin{figure}
\epsscale{0.63}
\plotone{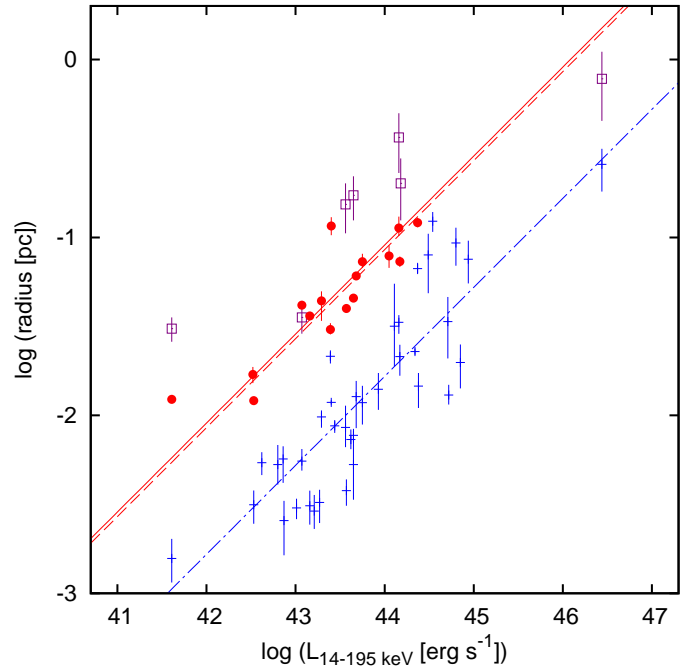}
\caption[Radii of the innermost dust torus and the BLR against the hard X-ray (14--195 keV) luminosity]{
Radii of the innermost dust torus and the broad emission-line region (BLR)
plotted against the hard X-ray (14--195 keV) luminosity.
Symbols are the same as Figure 13 but for the data
of the reverberation radii of broad Balmer emission lines
taken from more literature
\citep{bent+06b,bent+07,bent+09a,bent+09b,denn+10,bart+11a,bart+11b,grie+12b}.
}
\label{fig:dis:mvdt:lrxo}
\end{figure}

%

\clearpage

\setcounter{figure}{14}
\begin{figure}
\epsscale{0.63}
\plotone{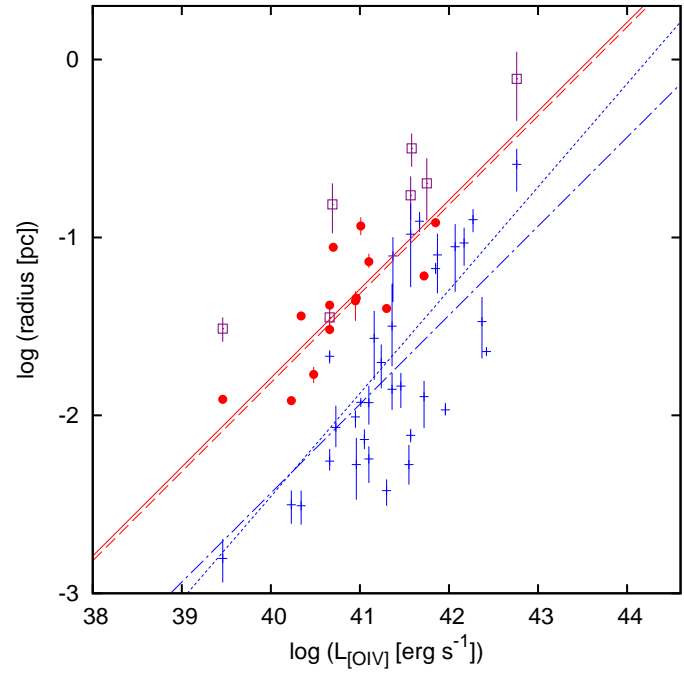}
\caption[Radii of the innermost dust torus and the BLR against the $\left[$\ion{O}{4} emission-line luminosity]{
Radii of the innermost dust torus and the broad emission-line region (BLR)
plotted against the [\ion{O}{4}] emission-line luminosity.
Symbols are the same as Figure 14 but
the dotted line (colored blue in the online version) represents
the best-fit regression line for the Balmer-line reverberation radii
presented by \citet{gree+10}.
}
\label{fig:dis:mvdt:lrxo_2}
\end{figure}

\clearpage


\begin{deluxetable}{lcccccccccc}
\tablewidth{0pt}
\tabletypesize{\footnotesize}
\tablecaption{List of target active galactic nuclei}
\tablehead{
\colhead{Name}     & \colhead{R.A.} & \colhead{Decl.} & \colhead{Diameter\tablenotemark{a}} &
\colhead{$z$} & \colhead{$M_{V}$\tablenotemark{b}} &
\colhead{Seyfert} & \colhead{Ref.} &
\colhead{$\log (M_{\rm{BH}}/M_{\odot})$} &
\colhead{Ref.} \\
\colhead{}         & \colhead{(J2000)} & \colhead{(J2000)} & \colhead{(arcmin)} &
\colhead{}   & \colhead{(mag)}          &
\colhead{type} & \colhead{} & \colhead{} & \colhead{}}
\startdata
  Mrk 335          & $00~06~19.5$&$+20~12~10.5$ & $0.3\times 0.3$ & $0.0258$ & $-22.10$ &
		       1.2 & 1 & $7.40\pm0.05$ & 3 \\
  Mrk 590          & $02~14~33.6$&$-00~46~00.1$ & $1.1\times 1.0$ & $0.0264$ & $-22.19$ &
                       1.2 & 2 & $7.68\pm0.07$ & 4 \\
  IRAS 03450$+$0055  & $03~47~40.2$&$+01~05~14.0$ & $0.28\times 0.27$ & $0.0306$ & $-21.67$ &
                       1   & \nodata  & $N/A$      & \nodata \\
  Akn 120          & $05~16~11.4$&$-00~08~59.4$ & $1.1\times 0.8$ & $0.0323$ & $-22.51$ &
                       1   & \nodata & $8.18\pm0.06$ & 4 \\
  MCG $+$08--11--011& $05~54~53.5$&$+46~26~22.0$ & $2.1\times 1.5$ & $0.0205$ & $-20.83$ &
                       1.5 & 2 & $N/A$         & \nodata \\
  Mrk 79           & $07~42~32.8$&$+49~48~34.8$ & $1.2\times 1.2$ & $0.0222$ & $-21.35$ &
                       1.2 & 2 & $7.72\pm0.12$ & 4 \\
  Mrk 110          & $09~25~12.9$&$+52~17~10.5$ & $0.67\times 0.42$ & $0.0353$ & $-21.26$ &
                       1.0 & 2 & $7.40\pm0.11$ & 4 \\
  NGC 3227         & $10~23~30.6$&$+19~51~54.0$ & $5.4\times 3.6$ & $0.0039$ & $-20.03$ &
                       1.5 & 2 & $6.88\pm0.10$ & 5 \\
  NGC 3516         & $11~06~47.5$&$+72~34~07.0$ & $1.7\times 1.3$ & $0.0082$ & $-21.07$ &
                       1.5 & 2 & $7.50\pm0.05$ & 5 \\
  Mrk 744          & $11~39~42.5$&$+31~54~33.0$ & $2.2\times 1.3$ & $0.0100$ & $-20.16$ &
                       1.8 & 2 & $N/A$         & \nodata \\
  NGC 4051         & $12~03~09.6$&$+44~31~52.8$ & $5.2\times 3.9$ & $0.0024$ & $-17.89$ &
                       1.5 & 1 & $6.24\pm0.13$ & 5 \\
  NGC 4151         & $12~10~32.6$&$+39~24~20.6$ & $6.3\times 4.5$ & $0.0033$ & $-19.64$ &
                       1.5 & 2 & $7.66\pm0.05$ & 6 \\
  NGC 4593         & $12~39~39.4$&$-05~20~39.4$ & $3.9\times 2.9$ & $0.0090$ & $-19.70$ &
                       1   & \nodata & $6.73\pm0.66$ & 4 \\
  NGC 5548         & $14~17~59.5$&$+25~08~12.4$ & $1.4\times 1.3$ & $0.0172$ & $-21.33$ &
                       1.5 & 2 & $7.82\pm0.02$ & 7 \\
  Mrk 817          & $14~36~22.1$&$+58~47~39.4$ & $0.6\times 0.6$ & $0.0315$ & $-22.59$ &
                       1.5 & 2 & $7.64\pm0.11$ & 5 \\
  Mrk 509          & $20~44~09.7$&$-10~43~24.5$ & $0.5\times 0.5$ & $0.0344$ & $-23.45$ &
                       1.2 & 2 & $8.16\pm0.04$ & 4 \\
  NGC 7469         & $23~03~15.6$&$+08~52~26.4$ & $1.5\times 1.1$ & $0.0163$ & $-21.91$ &
                       1.2 & 2 & $7.09\pm0.05$ & 4 
\enddata
\label{tab:obs:tar:stdinfo}
\tablerefs{(1) \citet{oste+93}; (2) \citet{vero+10};
 (3) \citet{grie+12b}; (4) \citet{pete+04}; (5) \citet{denn+10};
 (6) \citet{bent+06a}; (7) \citet{bent+07}.
}
\tablenotetext{a}{Major and minor diameters taken from the NED.}
\tablenotetext{b}{$V$-band absolute magnitude reported by \citet{vero+01}.}
\end{deluxetable}

\clearpage

\begin{deluxetable}{lccrr}
\tablewidth{0pt}
\tabletypesize{\footnotesize}
\tablecaption{Summary of monitoring parameters}
\tablehead{
\colhead{Object} & \colhead{Band} & \colhead{Observing period} &
\colhead{$n_{\rm{obs}}$\tablenotemark{a}} &
\colhead{$t_{\rm{int,obs}}$\tablenotemark{b}}}
\startdata
  Mrk 335        & $V$ & 2003.01.08---2006.01.18 &  41 & 15.0 \\
                 & $K$ & 2003.01.08---2006.01.18 &  40 & 15.0 \\
  Mrk 590        & $V$ & 2003.01.03---2007.08.08 &  92 &  6.5 \\
                 & $K$ & 2003.01.03---2007.08.08 &  94 &  6.0 \\
  IRAS 03450$+$0055  & $V$ & 2002.01.16---2007.08.19 &  88 &  9.8 \\
                 & $K$ & 2002.01.16---2007.08.19 &  86 &  9.8 \\
  Akn 120        & $V$ & 2001.11.24---2007.08.07 &  67 & 13.5 \\
                 & $K$ & 2001.11.24---2007.08.07 &  69 & 13.5 \\
  MCG $+$08--11--011       & $V$ & 2001.01.26---2007.08.19 & 178 &  4.0 \\
                 & $K$ & 2001.01.26---2007.08.19 & 187 &  4.0 \\
  Mrk 79         & $V$ & 2003.09.08---2006.05.21 &  47 & 10.0 \\
                 & $K$ & 2003.09.08---2006.05.11 &  47 & 10.0 \\
  Mrk 110        & $V$ & 2003.02.06---2006.05.26 &  49 & 15.8 \\
                 & $K$ & 2003.02.06---2006.05.26 &  50 & 15.0 \\
  NGC 3227       & $V$ & 2001.11.13---2007.07.01 & 123 &  6.0 \\
                 & $K$ & 2001.11.13---2007.07.01 & 132 &  6.0 \\
  NGC 3516       & $V$ & 2005.01.16---2007.06.26 &  35 & 14.0 \\
                 & $K$ & 2005.01.16---2007.07.02 &  38 & 13.1 \\
  Mrk 744        & $V$ & 2002.04.26---2006.05.25 &  90 &  7.0 \\
                 & $K$ & 2002.04.26---2006.05.25 &  94 &  7.9 \\
  NGC 4051       & $V$ & 2001.03.15---2007.07.30 & 158 &  7.0 \\
                 & $K$ & 2001.03.15---2007.07.30 & 181 &  6.0 \\
  NGC 4151       & $V$ & 2001.01.06---2007.08.07 & 234 &  4.9 \\
                 & $K$ & 2001.01.06---2007.08.07 & 239 &  4.9 \\
  NGC 4593       & $V$ & 2005.01.21---2007.07.09 &  35 & 13.0 \\
                 & $K$ & 2005.01.21---2007.07.27 &  37 & 13.4 \\
  NGC 5548       & $V$ & 2001.03.24---2007.08.20 & 302 &  3.0 \\
                 & $K$ & 2001.03.24---2007.08.20 & 311 &  3.0 \\
  Mrk 817        & $V$ & 2003.09.04---2006.05.17 &  38 & 16.0 \\
                 & $K$ & 2003.09.04---2006.05.17 &  40 & 16.0 \\
  Mrk 509        & $V$ & 2003.04.29---2006.05.21 &  45 & 13.1 \\
                 & $K$ & 2003.04.29---2006.05.21 &  49 & 13.5 \\
  NGC 7469       & $V$ & 2001.06.15---2008.12.08 & 296 &  4.0 \\
                 & $K$ & 2001.06.12---2008.12.08 & 315 &  4.1 
\enddata
\label{tab:obs:sch}
\tablenotetext{a}{Number of observations.}
\tablenotetext{b}{Median monitoring interval,
excluding lengthy observation gaps caused by
solar conjunction and occasional facility maintenance.}
\end{deluxetable}

\clearpage

\begin{deluxetable}{lcccrr}
\tablewidth{0pt}
\tabletypesize{\footnotesize}
\tablecaption{Magnitudes of reference stars}
\tablehead{
\colhead{Object} & \colhead{Reference} &
\colhead{R.A.} & \colhead{Decl.} &
\colhead{$m_V$} & \colhead{$m_K$}\\
\colhead{}       & \colhead{star} &
\colhead{(J2000)} & \colhead{(J2000)} &
\colhead{(mag)} & \colhead{(mag)}}
\startdata
        Mrk 335  & A & $00~07~04.9$&$+20~09~53.5$ & $12.859\pm0.004$ & $11.348\pm0.014$\\
                 & B & $00~05~51.5$&$+20~12~33.1$ & $13.443\pm0.004$ & $11.354\pm0.012$\\ 

        Mrk 590  & A1\tablenotemark{a}& $02~14~36.0$&$-00~39~11.0$ & $14.128\pm0.007$ & $13.474\pm0.010$\\
                 & A2\tablenotemark{a}& $02~14~32.2$&$-00~37~44.4$ & $12.899\pm0.005$ & $11.427\pm0.006$\\ 
                 & B & $02~14~35.1$&$-00~50~35.6$ & $14.092\pm0.004$ & $12.726\pm0.005$\\ 

        IRAS 03450$+$0055 & A & $03~47~17.7$&$+01~06~33.9$ & $12.865\pm0.004$ & $11.063\pm0.003$\\
                       & B & $03~47~43.5$&$+01~00~09.2$ & $13.208\pm0.005$ & $11.406\pm0.003$\\ 

        Akn 120  & A & $05~15~47.9$&$-00~07~26.3$ & $13.947\pm0.004$ & $11.138\pm0.006$\\
                 & B & $05~16~15.1$&$-00~11~45.2$ & $11.880\pm0.005$ & $10.342\pm0.005$\\ 

        MCG $+$08--11--011 & A & $05~55~07.6$&$+46~28~50.8$ & $13.752\pm0.002$ & $11.041\pm0.004$\\
                       & B & $05~54~47.8$&$+46~23~05.0$ & $12.520\pm0.002$ & $10.680\pm0.004$\\ 

        Mrk 79   & A & $07~42~08.2$&$+49~45~08.1$ & $13.402\pm0.005$ & $11.122\pm0.004$\\
                 & B & $07~42~52.2$&$+49~52~37.4$ & $12.465\pm0.004$ & $10.866\pm0.004$\\ 

        Mrk 110  & A & $09~24~57.1$&$+52~13~25.6$ & $14.151\pm0.007$ & $12.540\pm0.006$\\
                 & B & $09~26~45.9$&$+52~19~38.5$ & $14.422\pm0.006$ & $12.893\pm0.006$\\ 

        NGC 3227 & A & $10~23~13.5$&$+19~56~24.6$ & $12.775\pm0.002$ & $11.098\pm0.003$\\
                 & B & $10~24~02.9$&$+19~46~30.0$ & $11.963\pm0.003$ & $10.467\pm0.003$\\ 
 
        NGC 3516 & A & $11~03~46.0$&$+72~25~36.9$ & $12.846\pm0.008$ & $10.906\pm0.009$\\
                 & B & $11~08~33.2$&$+72~37~57.1$ & $13.429\pm0.009$ & $11.532\pm0.011$\\ 

        Mrk 744  & A & $11~38~59.2$&$+31~59~26.4$ & $14.259\pm0.007$ & $12.617\pm0.006$\\
                 & B & $11~39~58.2$&$+31~50~04.3$ & $14.114\pm0.006$ & $12.619\pm0.006$\\ 

        NGC 4051 & A & $12~04~13.1$&$+44~55~01.5$ & $12.643\pm0.003$ & $9.976\pm0.005$\\
                 & B & $12~03~30.1$&$+44~16~45.5$ & $13.881\pm0.003$ & $10.273\pm0.005$\\ 

        NGC 4151 & A & $12~11~37.2$&$+39~29~59.7$ & $11.926\pm0.002$ & $10.513\pm0.003$\\
                 & B & $12~09~47.3$&$+39~15~02.8$ & $11.913\pm0.002$ & $10.592\pm0.003$\\

        NGC 4593 & A & $12~40~08.0$&$-05~22~35.7$ & $12.711\pm0.003$ & $10.987\pm0.010$\\
                 & B & $12~39~16.6$&$-05~27~12.3$ & $13.109\pm0.004$ & $11.429\pm0.007$\\

        NGC 5548 & A & $14~17~58.8$&$+25~05~33.3$ & $13.789\pm0.002$ & $12.193\pm0.003$\\
                 & B & $14~17~14.7$&$+25~19~21.7$ & $13.134\pm0.002$ & $11.529\pm0.003$\\

        Mrk 817  & A & $14~36~00.7$&$+58~39~15.4$ & $13.437\pm0.006$ & $11.435\pm0.011$\\
                 & B & $14~36~26.8$&$+58~57~13.8$ & $12.807\pm0.006$ & $11.541\pm0.011$\\ 

        Mrk 509  & A & $20~44~48.5$&$-10~58~00.8$ & $13.402\pm0.005$ & $10.132\pm0.010$\\
                 & B & $20~43~39.1$&$-10~43~27.7$ & $12.646\pm0.004$ & $10.192\pm0.011$\\ 

        NGC 7469 & B\tablenotemark{b} & $23~02~23.3$&$+09~06~40.4$ & $12.999\pm0.001$ & $10.693\pm0.004$
\enddata
\label{tab:red:phot:abs:refstar1}
\tablenotetext{a}{The reference star was changed from A1 to A2 on July 12, 2004.}
\tablenotetext{b}{The reference star A for NGC 7469 was not used for photometry.}
\end{deluxetable}

\clearpage

\begin{deluxetable}{lcccc}
\tablewidth{0pt}
\tabletypesize{\footnotesize}
\tablecaption{Light curve data}
\tablehead{
\colhead{Object}        & \colhead{Band}  & \colhead{Observation Date}
                        & \colhead{Flux}  & \colhead{Flux Error} \\
\colhead{}              & \colhead{}      & \colhead{(MJD)}
                        & \colhead{(mJy)} & \colhead{(mJy)}}
\startdata
  Mrk 335               & $V$               & 52647.31
                        & 7.69              & 0.08 \\
Mrk 335               & $V$               & 52797.60
                        & 6.87              & 0.14 \\
\nodata               & \nodata           & \nodata
                        & \nodata           & \nodata \\
  Mrk 335               & $K$               & 52647.31
                        & 36.03             & 0.30 \\
Mrk 335               & $K$               & 52797.60
                        & 33.41             & 0.28 \\
\nodata               & \nodata           & \nodata
                        & \nodata           & \nodata
\enddata
\label{tab:lc}
\tablecomments{Correction for the Galactic extinction
has not been applied to the light curves, and
the fluxes from the host galaxy and narrow emission lines
have not been subtracted.
The complete data of this table is available in a machine-readable form
in the online journal. A portion is shown here for guidance.}
\end{deluxetable}

\clearpage

\begin{deluxetable}{lcccccc}
\tablewidth{0pt}
\tabletypesize{\footnotesize}
\tablecaption{Contribution of host-galaxy flux and narrow emission lines to the observed flux}
\tablehead{
\colhead{Object} & \colhead{Band} &
\colhead{$f_{\rm{host}}$\tablenotemark{a}} & \colhead{Ref.} &
\colhead{$f_{\rm{NL}}$\tablenotemark{b}} & \colhead{Ref.} &
\colhead{total} \\
\colhead{} & \colhead{} & \colhead{(mJy)} & \colhead{} &
\colhead{(mJy)} & \colhead{} &
\colhead{(mJy)}}
\startdata
  Mrk 335         & $V$ & $~2.03\pm0.10$ & 1 & $0.20$ & 1 & $~2.23\pm0.10$ \\
                  & $K$ & $~5.02\pm0.25$ & 1 & \nodata    &   & $~5.02\pm0.25$ \\
  Mrk 590         & $V$ & $~4.24\pm0.06$ & 2 & $0.16$ & 2 & $~4.39\pm0.06$ \\
                  & $K$ & $22.18\pm0.13$ & 1 & \nodata    &   & $22.18\pm0.13$ \\
  IRAS 03450$+$0055 & $V$ & $~0.65\pm0.09$ & 4 & $0.17$ & 1 & $~0.82\pm0.09$ \\
                  & $K$ & $~5.15\pm0.33$ & 4 & \nodata    &   & $~5.15\pm0.33$ \\
  Akn 120         & $V$ & $~3.09\pm0.07$ & 4 & $0.18$ & 1 & $~3.27\pm0.07$ \\
                  & $K$ & $20.02\pm0.77$ & 4 & \nodata    &   & $20.02\pm0.77$ \\
  MCG $+$08--11--011  & $V$ & $~2.24\pm0.17$ & 4 & $0.44$ & 1 & $~2.68\pm0.18$ \\
                  & $K$ & $16.78\pm1.00$ & 4 & \nodata    &   & $16.78\pm1.00$ \\
  Mrk 79          & $V$ & $~2.39\pm0.10$ & 1 & $0.24$ & 1 & $~2.63\pm0.11$ \\
                  & $K$ & $10.30\pm1.11$ & 1 & \nodata    &   & $10.30\pm1.11$ \\
  Mrk 110         & $V$ & $~0.71\pm0.09$ & 2 & $0.10$ & 2 & $~0.81\pm0.09$ \\
                  & $K$ & $~4.14\pm0.71$ & 1 & \nodata    &   & $~4.14\pm0.71$ \\
  NGC 3227        & $V$ & $~8.03\pm0.30$ & 2 & $0.51$ & 2 & $~8.53\pm0.30$ \\
                  & $K$ & $57.86\pm1.66$ & 3 & \nodata    &   & $57.86\pm1.66$ \\
  NGC 3516        & $V$ & $14.12\pm0.25$ & 2 & $0.21$ & 2 & $14.34\pm0.25$ \\
                  & $K$ & $68.81\pm0.53$ & 1 & \nodata    &   & $68.81\pm0.53$ \\
  Mrk 744         & $V$ & $~3.25\pm0.02$ & 1 & $0.08$ & 1 & $~3.33\pm0.02$ \\
                  & $K$ & $21.83\pm0.11$ & 1 & \nodata    &   & $21.83\pm0.11$ \\
  NGC 4051        & $V$ & $~7.74\pm0.36$ & 2 & $0.29$ & 2 & $~8.03\pm0.37$ \\
                  & $K$ & $39.05\pm1.36$ & 3 & \nodata    &   & $39.05\pm1.36$ \\
  NGC 4151        & $V$ & $17.17\pm0.83$ & 2 & $9.13$ & 2 & $26.30\pm1.23$ \\
                  & $K$ & $44.22\pm3.83$ & 5 & \nodata    &   & $44.22\pm3.83$ \\
  NGC 4593        & $V$ & $~6.73\pm0.08$ & 2 & $0.16$ & 2 & $~6.89\pm0.08$ \\
                  & $K$ & $34.40\pm0.22$ & 1 & \nodata    &   & $34.40\pm0.22$ \\
  NGC 5548        & $V$ & $~3.97\pm0.04$ & 2 & $0.37$ & 2 & $~4.34\pm0.05$ \\
                  & $K$ & $15.21\pm0.33$ & 3 & \nodata    &   & $15.21\pm0.33$ \\
  Mrk 817         & $V$ & $~1.26\pm0.04$ & 2 & $0.27$ & 2 & $~1.53\pm0.05$ \\
                  & $K$ & $~6.55\pm0.72$ & 1 & \nodata    &   & $~6.55\pm0.72$ \\
  Mrk 509         & $V$ & $~2.96\pm0.19$ & 1 & $0.82$ & 1 & $~3.78\pm0.21$ \\
                  & $K$ & $11.57\pm0.64$ & 1 & \nodata    &   &  $11.57\pm0.64$ \\
  NGC 7469        & $V$ & $~8.61\pm0.31$ & 3 & $0.69$ & 1 & $~9.30\pm0.31$ \\
                  & $K$ & $59.64\pm1.65$ & 3 & \nodata    &   & $59.64\pm1.65$
\enddata
\label{tab:hostflux}
\tablecomments{Correction for the Galactic extinction has not been applied.}
\tablenotetext{a}{The host-galaxy flux within the photometric aperture of $\phi = 8''.3$.}
\tablenotetext{b}{The contribution of [\ion{O}{3}]$\lambda 4959$ and
 $\lambda5007$ emission lines to the $V$-band flux.}
\tablerefs{(1) This study; (2) \citet{saka+10};
 (3) \citet{suga+06}; (4) \citet{tomi+06}; (5) \citet{mine+04}.
}
\end{deluxetable}

\clearpage

\begin{deluxetable}{lcc}
\tablewidth{0pt}
\tabletypesize{\footnotesize}
\tablecaption{Parameters of power-law fitting of the $V$-band structure function}
\tablehead{
\colhead{Object} & \colhead{$\alpha$} & \colhead{$\beta$}}
\startdata
    Mrk 335  & $0.0049$ & $0.88$ \\
    Mrk 590  & $0.0003$ & $1.02$ \\
    IRAS 03450$+$0055 & $0.0005$ & $1.38$ \\
    Akn 120  & $0.0385$ & $0.84$ \\
    MCG $+$08--11--011 & $0.0111$ & $0.94$ \\
    Mrk 79   & $0.0015$ & $1.34$ \\
    Mrk 110  & $0.0024$ & $1.21$ \\
    NGC 3227 & $0.0523$ & $0.67$ \\
    NGC 3516 & $0.0026$ & $1.83$ \\
    Mrk 744  & $0.0007$ & $1.03$ \\
    NGC 4051 & $0.0343$ & $0.78$ \\
    NGC 4151 & $0.7197$ & $0.92$ \\
    NGC 4593 & $0.0012$ & $1.44$ \\
    NGC 5548 & $0.0018$ & $1.46$ \\
    Mrk 817  & $0.0006$ & $1.39$ \\
    Mrk 509  & $0.0296$ & $1.02$ \\
    NGC 7469 & $0.0334$ & $0.87$
\enddata
\label{tab:ana:lc:sf}
\end{deluxetable}

\clearpage
\begin{deluxetable}{lcccrrrr}
\tablewidth{0pt}
\tabletypesize{\footnotesize}
\tablecaption{Lag time and $V$-band absolute magnitude for the targets}
\tablehead{
\colhead{Object} & \colhead{Section}\tablenotemark{a} & \colhead{MJD} &
\colhead{$M_{V}$} &
\colhead{$\Delta t_K({\rm CCF})$\tablenotemark{b}} &
\colhead{$\Delta t_K({\rm CCF})$\tablenotemark{c}} &
\colhead{$\Delta t_K({\rm JAVELIN})$\tablenotemark{b}} &
\colhead{$\Delta t_K({\rm JAVELIN})$\tablenotemark{c}} \\
\colhead{} & \colhead{} & \colhead{} &
\colhead{(mag)} &
\colhead{(days)} &
\colhead{(days)} &
\colhead{(days)} & 
\colhead{(days)}}
\startdata
Mrk 335     
  & 1 & 52647.3---53753.3 & $-20.63\pm 0.02$ &
                            $142.2_{-16.6}^{+16.8}$ &
                            $132.1_{-13.6}^{+18.0}$ &
                            $167.1_{~-5.9}^{~+5.4}$ &
                            $165.8_{~-7.4}^{~+6.0}$ \\
  & WA        & \nodata   & $-20.63\pm 0.02$ &
                            $142.2\pm16.7$ &
                            $132.1\pm15.8$ &
                            $167.1\pm 5.6$ &
                            $165.8\pm 6.7$ \\
Mrk 590      
  & 1 & 52842.6---53429.2 & $-18.69\pm 0.08$ &
                            $~37.2_{~-2.3}^{~+3.1}$ &
                            $~35.9_{~-2.7}^{~+2.7}$ &
                            $~33.8_{~-4.2}^{~+4.2}$ &
                            $~32.2_{~-4.0}^{~+4.0}$ \\
  & WA        & \nodata   & $-18.69\pm 0.08$ &
                            $~37.2\pm2.7$ &
                            $~35.9\pm2.7$ &
                            $~33.8\pm 4.2$ &
                            $~32.2\pm 4.0$ \\
IRAS 03450$+$0055     
  & 1 & 52290.4---52709.3 & $-21.48\pm 0.02$ &
                            $~95.5_{~-4.4}^{~+7.3}$ &
                            $~93.1_{~-4.8}^{~+6.0}$ &
                            $156.3_{~-8.4}^{+10.7}$ &
                            $136.8_{-13.5}^{+14.9}$ \\
  & 2 & 52574.5---53453.2 & $-21.30\pm 0.03$ &
                            $167.3_{-11.7}^{+10.2}$ &
                            $146.0_{-18.2}^{+14.0}$ &
                            $166.3_{~-6.8}^{~+6.8}$ &
                            $165.3_{~-9.4}^{~+8.0}$ \\
  & 3 & 53578.6---54013.4 & $-21.53\pm 0.02$ &
                            $~88.5_{~-8.8}^{+16.4}$ &
                            $~70.3_{~-8.1}^{+10.7}$ &
                            $~97.5_{-15.9}^{+24.3}$ &
                            $~95.0_{-22.4}^{+27.9}$ \\
  & WA        & \nodata   & $-21.41\pm 0.02$ &
                            $108.2\pm4.8$ &
                            $~92.0\pm4.5$ &
                            $158.3\pm 5.4$ &
                            $152.5\pm 7.1$ \\
Akn 120     
  & 1 & 53594.6---54319.6 & $-22.23\pm 0.02$ &
                            $138.8_{-12.8}^{+22.1}$ &
                            $127.6_{-13.1}^{+17.2}$ &
                            $140.9_{-15.9}^{+18.5}$ &
                            $129.7_{-16.8}^{+19.3}$ \\
  & WA        & \nodata   & $-22.23\pm 0.02$ &
                            $138.8\pm 17.5$ &
                            $127.6\pm 15.2$ &
                            $140.9\pm 17.2$ &
                            $129.7\pm 18.0$ \\
MCG $+$08$-$11$-$011     
  & 1 & 52158.6---53124.3 & $-20.62\pm 0.04$ &
                            $111.1_{-21.3}^{+20.0}$ &
                            $114.9_{-22.0}^{+18.3}$ &
                            $~63.9_{~-1.3}^{~+1.7}$ &
                            $~69.3_{~-2.1}^{~+2.5}$ \\
  & 2 & 53593.6---54224.3 & $-19.27\pm 0.16$ &
                            $~83.4_{-23.2}^{+14.1}$ &
                            $~67.6_{-21.3}^{+11.5}$ &
           \tablenotemark{d}$127.1_{~-3.6}^{~+3.6}$ &
                            $~72.1_{~-6.5}^{~+2.7}$ \\
  & WA        & \nodata   & $-20.25\pm 0.06$ &
                            $~95.8\pm13.8$ &
                            $~86.4\pm12.7$ &
                            $~73.5\pm1.4$ &
                            $~69.9\pm2.1$ \\
Mrk 79     
  & 1 & 52890.6---53150.3 & $-20.06\pm 0.04$ &
                            $~52.9_{~-6.2}^{~+8.8}$ &
                            $~50.7_{~-5.0}^{~+5.4}$ &
                            $~60.2_{~-6.0}^{~+5.9}$ &
                            $~57.9_{~-7.4}^{~+6.3}$ \\
  & 2 & 53369.3---53876.3 & $-19.78\pm 0.05$ &
                            $~80.3_{~-4.6}^{~+3.6}$ &
                            $~72.8_{~-3.0}^{~+3.1}$ &
                            $~78.6_{~-9.5}^{~+8.9}$ &
                            $~80.4_{~-5.5}^{~+5.3}$ \\
  & WA        & \nodata   & $-19.98\pm 0.04$ &
                            $~74.0\pm3.6$ &
                            $~67.0\pm2.6$ &
                            $~65.6\pm5.0$ &
                            $~71.8\pm4.2$ \\
Mrk 110     
  & 1 & 52676.3---53163.3 & $-19.90\pm 0.09$ &
                            $~87.6_{~-8.5}^{~+7.6}$ &
                            $~80.2_{~-8.5}^{~+7.1}$ &
                            $113.1_{~-8.6}^{~+8.8}$ &
                            $103.2_{-13.5}^{+10.9}$ \\
  & 2 & 53291.6---53881.3 & $-20.94\pm 0.03$ &
                            $~93.8_{-10.5}^{~+8.8}$ &
                            $~89.3_{~-9.2}^{~+7.9}$ &
           \tablenotemark{d}$124.1_{~-7.1}^{~+7.1}$ &
                            $109.4_{~-8.8}^{+12.3}$ \\
  & WA        & \nodata   & $-20.63\pm 0.04$ &
                            $~90.3\pm6.1$ &
                            $~84.3\pm5.8$ &
                            $119.7\pm5.5$ &
                            $106.8\pm8.0$ \\
NGC 3227     
  & 1 & 52226.6---52423.3 & $-16.88\pm 0.08$ &
                            $~18.9_{~-4.8}^{~+4.3}$ &
                            $~18.2_{~-4.9}^{~+4.4}$ &
                            $~26.9_{~-3.8}^{~+3.9}$ &
                            $~25.3_{~-3.8}^{~+4.3}$ \\
  & 2 & 52935.6---53176.3 & $-16.95\pm 0.07$ &
                            $~17.4_{~-4.4}^{~+3.6}$ &
                            $~15.6_{~-3.7}^{~+3.2}$ &
                            $~15.8_{~-2.0}^{~+2.4}$ &
                            $~14.9_{~-1.9}^{~+2.2}$ \\
  & 3 & 53350.5---53485.4 & $-17.25\pm 0.05$ &
                            $~19.7_{~-1.5}^{~+1.3}$ &
                            $~18.9_{~-1.3}^{~+1.4}$ &
                            $~16.9_{~-4.7}^{+10.8}$ &
                            $~14.0_{~-2.3}^{~+3.2}$ \\
  & 4 & 53664.6---53790.4 & $-17.07\pm 0.06$ &
                            $~10.7_{~-1.6}^{~+1.7}$ &
                            $~10.0_{~-1.7}^{~+1.9}$ &
                            $~10.2_{~-2.5}^{~+2.8}$ &
                            $~~9.4_{~-2.6}^{~+3.0}$ \\
  & 5 & 53723.6---53904.3 & $-16.96\pm 0.07$ &
                            $~15.4_{~-1.8}^{~+1.7}$ &
                            $~14.8_{~-1.9}^{~+1.7}$ &
                            $~18.7_{~-3.0}^{~+5.6}$ &
                            $~19.6_{~-3.6}^{~+4.5}$ \\
  & 6 & 54181.4---54282.3 & $-16.99\pm 0.06$ &
                            $~12.9_{~-0.8}^{~+1.0}$ &
                            $~11.6_{~-0.8}^{~+0.8}$ &
                            $~14.1_{~-0.8}^{~+0.8}$ &
                            $~13.2_{~-0.8}^{~+0.9}$ \\
  & WA        & \nodata   & $-17.06\pm 0.06$ &
                            $~14.5\pm0.6$ &
                            $~13.3\pm0.6$ &
                            $~14.6\pm0.7$ &
                            $~13.8\pm0.7$ \\
NGC 3516     
  & 1 & 53386.4---53920.3 & $-18.14\pm 0.09$ &
                            $~52.9_{-12.0}^{~+7.0}$ &
                            $~47.3_{-11.1}^{~+8.7}$ &
                            $~73.1_{~-4.0}^{~+3.9}$ &
                            $~71.5_{~-6.2}^{~+5.4}$ \\
  & WA        & \nodata   & $-18.14\pm 0.09$ &
                            $~52.9\pm 9.5$ &
                            $~47.3\pm 9.9$ &
                            $~73.1\pm 4.0$ &
                            $~71.5\pm 5.8$ \\
Mrk 744     
  & 1 & 52994.5---53208.3 & $-16.18\pm 0.06$ &
                            $~16.2_{~-2.7}^{~+2.4}$ &
                            $~15.6_{~-2.2}^{~+2.3}$ &
                            $~17.0_{~-3.1}^{~+3.1}$ &
                            $~16.6_{~-3.2}^{~+3.1}$ \\
  & 2 & 53336.6---53550.3 & $-16.04\pm 0.08$ &
                            $~28.5_{~-3.5}^{~+3.7}$ &
                            $~28.1_{~-3.2}^{~+3.9}$ &
                            $~23.0_{~-2.8}^{~+3.3}$ &
                            $~22.6_{~-2.9}^{~+3.3}$ \\
  & WA        & \nodata   & $-16.12\pm 0.05$ &
                            $~20.4\pm2.1$ &
                            $~19.2\pm1.9$ &
                            $~20.0\pm2.2$ &
                            $~19.7\pm2.2$ \\
NGC 4051      
  & 1 & 52286.6---52488.3 & $-15.74\pm 0.10$ &
                            $~12.5_{~-1.1}^{~+1.0}$ &
                            $~11.9_{~-1.2}^{~+1.0}$ &
           \tablenotemark{d}$~15.9_{~-1.7}^{~+1.7}$ &
                            $~14.5_{~-0.8}^{~+0.9}$ \\
  & 2 & 52599.6---52832.3 & $-15.78\pm 0.09$ &
                            $~23.4_{~-2.4}^{~+2.4}$ &
                            $~22.9_{~-2.3}^{~+2.3}$ &
                            $~24.8_{~-1.0}^{~+1.1}$ &
                            $~25.0_{~-1.2}^{~+1.1}$ \\
  & 3 & 53341.6---53577.3 & $-15.69\pm 0.10$ &
                            $~20.2_{~-5.5}^{~+4.6}$ &
                            $~18.3_{~-4.4}^{~+4.6}$ &
           \tablenotemark{d}$~24.0_{-12.9}^{+12.9}$ &
           \tablenotemark{d}$~22.9_{-13.4}^{+13.4}$ \\
  & 4 & 53684.6---53937.3 & $-15.73\pm 0.10$ &
                            $~16.3_{~-1.2}^{~+1.3}$ &
                            $~15.5_{~-1.1}^{~+1.2}$ &
                            $~11.9_{~-0.9}^{~+1.2}$ &
                            $~13.4_{~-1.4}^{~+1.0}$ \\
  & 5 & 54202.5---54252.4 & $-15.77\pm 0.09$ &
                            $~11.0_{~-1.6}^{~+0.9}$ &
                            $~10.7_{~-1.1}^{~+1.4}$ &
                            $~~9.9_{~-1.0}^{~+2.5}$ &
                            $~~9.6_{~-0.9}^{~+2.3}$ \\
  & 6 & 54221.5---54278.3 & $-15.75\pm 0.09$ &
                            $~12.4_{~-1.1}^{~+1.2}$ &
                            $~12.3_{~-1.1}^{~+1.2}$ &
                            $~13.0_{~-2.4}^{~+1.9}$ &
                            $~12.7_{~-2.6}^{~+2.1}$ \\
  & 7 & 54252.4---54311.3 & $-15.73\pm 0.10$ &
                            $~16.7_{~-0.7}^{~+0.8}$ &
                            $~16.2_{~-0.8}^{~+0.7}$ &
                            $~15.2_{~-2.0}^{~+2.1}$ &
                            $~14.6_{~-2.1}^{~+1.9}$ \\
  & WA        & \nodata   & $-15.75\pm 0.09$ &
                            $~14.7\pm0.5$ &
                            $~14.3\pm0.4$ &
                            $~16.5\pm0.6$ &
                            $~15.8\pm0.5$ \\
NGC 4151      
  & 1 & 51915.6---52127.3 & $-17.78\pm 0.13$ &
                            $~59.6_{~-1.3}^{~+1.1}$ &
                            $~53.9_{~-2.4}^{~+1.7}$ &
                            $~59.7_{~-1.4}^{~+1.5}$ &
                            $~56.6_{~-1.2}^{~+1.1}$ \\
  & 2 & 52265.5---52646.6 & $-18.59\pm 0.07$ &
                            $~70.3_{-13.0}^{+14.0}$ &
                            $~64.7_{-16.3}^{+19.6}$ &
                            $~68.8_{~-1.3}^{~+2.0}$ &
                            $~70.0_{~-2.7}^{~+4.4}$ \\
  & 3 & 52610.6---53010.5 & $-18.95\pm 0.05$ &
                            $~59.1_{~-2.4}^{~+2.0}$ &
                            $~56.4_{~-3.1}^{~+2.1}$ &
                            $~60.1_{~-1.5}^{~+1.8}$ &
                            $~58.1_{~-1.6}^{~+1.8}$ \\
  & 4 & 53039.4---53356.5 & $-18.19\pm 0.10$ &
                            $~65.1_{~-3.0}^{~+2.9}$ &
                            $~62.9_{~-3.0}^{~+2.8}$ &
                            $~66.2_{~-2.9}^{~+8.1}$ &
                            $~65.3_{~-2.8}^{~+3.1}$ \\
  & 5 & 53148.4---53490.3 & $-18.09\pm 0.11$ &
                            $~50.4_{-12.1}^{~+6.9}$ &
                            $~45.1_{-12.0}^{~+7.1}$ &
                            $~81.2_{~-2.2}^{~+3.1}$ &
                            $~82.7_{~-3.0}^{~+4.2}$ \\
  & 6 & 53436.4---53687.6 & $-17.94\pm 0.13$ &
                            $~41.2_{~-3.8}^{~+5.1}$ &
                            $~37.7_{~-3.3}^{~+4.8}$ &
                            $42.0_{~-0.8}^{~+0.4}$ &
                            $41.9_{~-1.3}^{~+0.5}$ \\
  & 7 & 53687.6---53864.4 & $-18.21\pm 0.09$ &
                            $~33.1_{~-2.3}^{~+2.3}$ &
                            $~31.8_{~-2.7}^{~+1.9}$ &
           \tablenotemark{d}$38.7_{~-2.6}^{~+2.6}$ &
                            $~36.0_{~-0.4}^{~+0.5}$ \\
  & 8 & 53753.4---53921.3 & $-18.02\pm 0.10$ &
                            $~38.5_{~-1.2}^{~+1.3}$ &
                            $~36.8_{~-1.2}^{~+1.1}$ &
                            $~41.7_{~-1.0}^{~+1.9}$ &
                            $~40.2_{~-0.7}^{~+0.6}$ \\
  & WA        & \nodata   & $-18.14\pm 0.09$ &
                            $~49.7\pm0.7$ &
                            $~43.1\pm0.8$ &
                            $~48.3\pm0.5$ &
                            $~40.9\pm0.3$ \\
NGC 4593     
  & 1 & 53391.5---53930.3 & $-17.93\pm 0.04$ &
                            $~43.5_{~-1.6}^{~+1.9}$ &
                            $~44.0_{~-1.8}^{~+2.1}$ &
                            $~41.6_{~-0.8}^{~+0.9}$ &
                            $~42.1_{~-1.1}^{~+0.8}$ \\
  & WA        & \nodata   & $-17.93\pm 0.04$ &
                            $43.5\pm 1.8$ &
                            $44.0\pm 2.0$ &
                            $41.6\pm 0.9$ &
                            $42.1\pm 0.9$ \\
NGC 5548      
  & 1 & 51992.5---52389.5 & $-19.33\pm 0.03$ &
                            $~62.9_{~-1.5}^{~+1.2}$ &
                            $~57.4_{~-1.4}^{~+1.4}$ &
                            $~63.8_{~-0.9}^{~+0.8}$ &
                            $~56.8_{~-2.6}^{~+1.7}$ \\
  & 2 & 52308.6---52797.4 & $-18.84\pm 0.05$ &
                            $~67.7_{~-5.7}^{~+5.4}$ &
                            $~61.5_{~-5.7}^{~+4.7}$ &
                            $~59.9_{~-0.4}^{~+0.4}$ &
                            $~55.3_{~-0.4}^{~+1.1}$ \\
  & 3 & 52638.6---52999.6 & $-19.13\pm 0.04$ &
                            $~62.5_{~-4.1}^{~+3.2}$ &
                            $~59.3_{~-3.8}^{~+3.1}$ &
                            $~79.3_{~-1.9}^{~+1.7}$ &
                            $~76.3_{~-2.0}^{~+1.8}$ \\
  & 4 & 53168.4---53437.5 & $-18.57\pm 0.05$ &
                            $~49.1_{~-5.2}^{~+4.8}$ &
                            $~45.7_{~-4.8}^{~+4.3}$ &
           \tablenotemark{d}$~45.5_{~-5.0}^{~+5.0}$ &
           \tablenotemark{d}$~44.1_{~-5.9}^{~+5.9}$ \\
  & 5 & 53350.6---53527.4 & $-18.56\pm 0.05$ &
                            $~41.7_{~-1.4}^{~+1.4}$ &
                            $~40.5_{~-1.4}^{~+1.5}$ &
                            $~42.8_{~-3.3}^{~+3.1}$ &
                            $~41.7_{~-3.1}^{~+2.9}$ \\
  & 6 & 54181.6---54332.3 & $-18.41\pm 0.05$ &
                            $~57.0_{~-0.9}^{~+1.0}$ &
                            $~55.1_{~-0.8}^{~+0.8}$ &
                            $~57.4_{~-1.9}^{~+1.9}$ &
                            $~62.7_{~-0.3}^{~+0.4}$ \\
  & WA        & \nodata   & $-18.77\pm 0.02$ &
                            $55.2\pm 0.7$ &
                            $53.0\pm 0.6$ &
                            $60.9\pm 0.3$ &
                            $61.3\pm 0.3$ \\
Mrk 817     
  & 1 & 53353.6---53872.5 & $-20.86\pm 0.02$ &
                            $~89.8_{~-7.0}^{~+9.6}$ &
                            $~83.5_{~-6.5}^{~+8.6}$ &
                            $~93.0_{~-8.5}^{~+9.4}$ &
                            $~89.8_{~-8.4}^{~+8.8}$ \\
  & WA        & \nodata   & $-20.86\pm 0.02$ &
                            $~89.8\pm 8.3$ &
                            $~83.5\pm 7.6$ &
                            $~93.0\pm 8.9$ &
                            $~89.8\pm 8.6$ \\
Mrk 509     
  & 1 & 52758.6---53867.6 & $-22.13\pm 0.03$ &
                            $149.2_{~-8.6}^{~+9.8}$ &
                            $133.4_{~-8.0}^{~+9.5}$ &
                            $120.7_{~-1.0}^{~+2.6}$ &
                            $120.3_{~-1.0}^{~+1.3}$ \\
  & WA        & \nodata   & $-22.13\pm 0.03$ &
                            $149.2\pm 9.2$ &
                            $133.4\pm 8.8$ &
                            $120.7\pm 1.8$ &
                            $120.3\pm 1.1$ \\
\enddata
\label{tab:dis:dtvar:tab2}
\end{deluxetable}

\clearpage

\begin{deluxetable}{lcccrrrr}
\setcounter{table}{7}
\tablewidth{0pt}
\tabletypesize{\footnotesize}
\tablecaption{(continued)}
\tablehead{
\colhead{Object} & \colhead{Section}\tablenotemark{a} & \colhead{MJD} &
\colhead{$M_{V}$} &
\colhead{$\Delta t_K({\rm CCF})$\tablenotemark{b}} &
\colhead{$\Delta t_K({\rm CCF})$\tablenotemark{c}} &
\colhead{$\Delta t_K({\rm JAVELIN})$\tablenotemark{b}} &
\colhead{$\Delta t_K({\rm JAVELIN})$\tablenotemark{c}} \\
\colhead{} & \colhead{} & \colhead{} &
\colhead{(mag)} &
\colhead{(days)} &
\colhead{(days)} &
\colhead{(days)} & 
\colhead{(days)}}
\startdata
NGC 7469     
  & 1 & 52072.5---52647.2 & $-19.85\pm 0.06$ &
                            $~97.5_{~-4.9}^{~+5.2}$ &
                            $~98.7_{~-5.7}^{~+5.8}$ &
                            $~88.4_{~-1.5}^{~+1.4}$ &
                            $~94.1_{~-0.5}^{~+0.2}$ \\
  & 2 & 52787.6---53369.3 & $-19.68\pm 0.07$ &
                            $~37.0_{~-2.8}^{~+2.1}$ &
                            $~37.9_{~-3.1}^{~+3.1}$ &
                            $106.8_{~-1.1}^{~+1.0}$ &
                            $106.4_{~-1.2}^{~+0.9}$ \\
  & 3 & 53508.6---54017.3 & $-19.84\pm 0.06$ &
                            $103.9_{~-4.0}^{~+4.2}$ &
                            $~92.5_{~-4.0}^{~+6.1}$ &
           \tablenotemark{d}$125.4_{~-2.7}^{~+2.7}$ &
                            $122.9_{~-0.2}^{~+0.2}$ \\
  & 4 & 54233.6---54808.3 & $-19.75\pm 0.07$ &
                            $~33.6_{~-2.6}^{~+1.5}$ &
                            $~32.4_{~-3.4}^{~+2.1}$ &
                            $~74.4_{~-1.5}^{~+0.1}$ &
                            $74.43_{-0.06}^{+0.05}$ \\
  & WA        & \nodata   & $-19.79\pm 0.06$ &
                            $~48.3\pm 1.4$ &
                            $~48.6\pm 1.8$ &
                            $~88.0\pm 0.6$ &
                            $78.12\pm 0.05$
\enddata
\tablecomments{The lag times in the observed frame were presented.}
\tablenotetext{a}{The row of the notation WA presents 
the weighted averaged $V$-band absolute magnitude
and the weighted averaged lag time
for the individual targets.}
\tablenotetext{b}{Assuming $\alpha_{\nu}=0$ for the subtraction of
the accretion-disk component in the $K$-band flux.}
\tablenotetext{c}{Assuming $\alpha_{\nu}=+1/3$ for the subtraction of
the accretion-disk component in the $K$-band flux.}
\tablenotetext{d}{The middle point of the 16 and 84 \%-ile lag times
of the likelihood distribution was adopted
because ambiguous multiple peaks appeared.}
\end{deluxetable}

\clearpage

\begin{deluxetable}{ccccccccc}
\tablewidth{0pt}
\tabletypesize{\footnotesize}
\tablecaption{Results of linear regression of dust lag-luminosity correlation}
\tablehead{
\colhead{Lag Method} & \colhead{N\tablenotemark{a}} & \colhead{$V_{\rm pec}$} &
\colhead{$\sigma_{\rm{add}}$\tablenotemark{c}} & \colhead{$a$\tablenotemark{c}} & \colhead{reduced $\chi^{2}$ \tablenotemark{c}} &
\colhead{$\sigma_{\rm{add}}$\tablenotemark{d}} & \colhead{$a$\tablenotemark{d}} & \colhead{reduced $\chi^{2}$ \tablenotemark{d}} \\
\colhead{} & \colhead{(km s$^{-1}$)} & \colhead{} & \colhead{} & \colhead{} & \colhead{} & \colhead{} & \colhead{} }
\startdata
CCF & 49 & ---  &  ---    & $-2.052\pm 0.004$ & $22.7$ &  ---    & $-2.074\pm 0.004$ & $21.2$ \\
    & 17 & ---  &  ---    & $-2.080\pm 0.005$ & $37.2$ &  ---    & $-2.093\pm 0.005$ & $38.0$ \\
    & 17 & 200  &  ---    & $-2.141\pm 0.008$ & $14.4$ &  ---    & $-2.166\pm 0.008$ & $15.9$ \\
    & 17 & 300  &  ---    & $-2.157\pm 0.010$ & $9.4$  &  ---    & $-2.187\pm 0.010$ & $10.5$ \\
    & 17 & ---  & $0.140$ & $-2.113\pm 0.035$ & $1.0$\tablenotemark{b} & $0.144$ & $-2.146\pm 0.036$ & $1.0$\tablenotemark{b}\\
    & 17 & 200  & $0.136$ & $-2.117\pm 0.035$ & $1.0$\tablenotemark{b} & $0.140$ & $-2.150\pm 0.036$ & $1.0$\tablenotemark{b}\\
    & 17 & 300  & $0.131$ & $-2.122\pm 0.035$ & $1.0$\tablenotemark{b} & $0.135$ & $-2.154\pm 0.036$ & $1.0$\tablenotemark{b}\\
JAVELIN  & 49 & ---  &  ---    & $-2.051\pm 0.003$ & $66.5$ &  ---    & $-2.057\pm 0.003$ & $72.5$ \\
    & 17 & ---  &  ---    & $-2.051\pm 0.003$ & $131.8$ &  ---    & $-2.072\pm 0.003$ & $157.0$ \\
    & 17 & 200  &  ---    & $-2.127\pm 0.006$ & $49.9$ &  ---    & $-2.157\pm 0.006$ & $48.5$ \\
    & 17 & 300  &  ---    & $-2.131\pm 0.008$ & $29.9$ &  ---    & $-2.156\pm 0.008$ & $28.2$ \\
    & 17 & ---  & $0.162$ & $-2.078\pm 0.040$ & $1.0$\tablenotemark{b} & $0.162$ & $-2.096\pm 0.040$ & $1.0$\tablenotemark{b}\\
    & 17 & 200  & $0.159$ & $-2.081\pm 0.040$ & $1.0$\tablenotemark{b} & $0.159$ & $-2.098\pm 0.040$ & $1.0$\tablenotemark{b}\\
    & 17 & 300  & $0.156$ & $-2.084\pm 0.040$ & $1.0$\tablenotemark{b} & $0.156$ & $-2.100\pm 0.040$ & $1.0$\tablenotemark{b}
\label{tab:mvdtfit}
\enddata

\tablecomments{The fitted model is $\log \Delta t_K=a-0.2 M_V$.}
\tablenotetext{a}{The number of the data pair for the fitting.
$N=49$ indicates that all measured lag times and absolute magnitudes
were used; $N=17$ indicates that the weighted averaged lag times
and absolute magnitudes for individual targets were used.}
\tablenotetext{b}{$\sigma_{\rm{add}}$ was added to the error of $\log \Delta t_K$
by root-sum-square for the reduced $\chi^{2}$ to achieve unity.}
\tablenotetext{c}{Assuming $\alpha_{\nu}=0$ for the subtraction of
the accretion-disk component in the $K$-band flux.}
\tablenotetext{d}{Assuming $\alpha_{\nu}=+1/3$ for the subtraction of
the accretion-disk component in the $K$-band flux.}

\end{deluxetable}

\clearpage

\begin{deluxetable}{ccccccl}
\tablewidth{0pt}
\tabletypesize{\footnotesize}
\tablecaption{Results of linear regression of radius-luminosity correlation}
\tablehead{
\colhead{Source} & \colhead{$L$} & \colhead{N\tablenotemark{a}} &
\colhead{$\alpha $} & \colhead{$\beta $\tablenotemark{b}} &
\colhead{$\sigma_{\rm{add}}$\tablenotemark{c}} & \colhead{Ref.}\\
\colhead{} & \colhead{(erg s$^{-1}$)} & \colhead{} & \colhead{} & 
\colhead{} & \colhead{} & \colhead{}}
\startdata
near-infrared interferometry & $L_V/10^{44}$     &  $9$ & $-0.72\pm 0.06$ & $0.5$ & 0.15 &  1\\ 
                             & $L_{\rm BAT}/10^{44}$   &  $6$ & $-0.63\pm 0.10$ & $0.5$ & 0.21 &  1\\ 
                             & $L_{\rm [OIV]}/10^{41}$ &  $7$ & $-0.94\pm 0.09$ & $0.5$ & 0.19 &  1\\ 
dust reverberation\tablenotemark{d}
                             & $L_V/10^{44}$     & $17$ & $-0.89\pm 0.04$ & $0.5$ & 0.14 &  1\\
                             &                         &      & $-0.92\pm 0.04$ & $0.5$ & 0.14 &  1\\
                             &                         &      & $-0.85\pm 0.04$ & $0.5$ & 0.16 &  1\\
                             &                         &      & $-0.87\pm 0.04$ & $0.5$ & 0.16 &  1\\
                             & $L_{\rm BAT}/10^{44}$   & $16$ & $-1.04\pm 0.04$ & $0.5$ & 0.16 &  1\\
                             &                         &      & $-1.07\pm 0.04$ & $0.5$ & 0.16 &  1\\
                             &                         &      & $-1.01\pm 0.05$ & $0.5$ & 0.19 &  1\\
                             &                         &      & $-1.03\pm 0.05$ & $0.5$ & 0.19 &  1\\
                             & $L_{\rm [OIV]}/10^{41}$ & $14$ & $-1.29\pm 0.06$ & $0.5$ & 0.21 &  1\\
                             &                         &      & $-1.32\pm 0.06$ & $0.5$ & 0.21 &  1\\
                             &                         &      & $-1.25\pm 0.07$ & $0.5$ & 0.25 &  1\\
                             &                         &      & $-1.26\pm 0.07$ & $0.5$ & 0.24 &  1\\
BLR reverberation            & $L_V/10^{44}$     & $34$ & $-1.51$	& $0.519^{+0.063}_{-0.066}$ & --- & 2\\
                             & $L_{\rm BAT}/10^{44}$   & $36$ & $-1.78\pm 0.04$ & $0.5$ & 0.25 &  1\\ 
                             & $L_{\rm [OIV]}/10^{41}$ & $33$ & $-1.94\pm 0.07$ & $0.5$ & 0.36 &  1\\ 
                             & $L_{\rm [OIV]}/10^{41}$ & $26$ & $-1.88\pm 0.10$ & $0.58\pm 0.11$ & 0.35 & 3
\enddata
\label{tab:rlfit}
\tablecomments{The fitted model is $\log r=\alpha + \beta \log L$.}
\tablerefs{(1) This study; (2) \citet{bent+09b}; (3) \citet{gree+10}.}
\tablenotetext{a}{The number of the data pair for the fitting.}
\tablenotetext{b}{$\beta =0.5$ is assumed for the fitting when that value is presented.}
\tablenotetext{c}{$\sigma_{\rm{add}}$ was added to the error of $\log r$
by root-sum-square for the reduced $\chi^{2}$ to achieve unity.}
\tablenotetext{d}{The lag times $\Delta t$ were derived
by the two different lag-analysis methods assuming the two different $\alpha_{\nu}$ values
for the subtraction of the accretion-disk component in the $K$-band flux.
The values of $\alpha $ and $\sigma_{\rm{add}}$ in four rows for each lag-luminosity correlation
were obtained from the lag times of
$\Delta t({\rm CCF};\ \alpha_{\nu}=0)$,
$\Delta t({\rm CCF};\ \alpha_{\nu}=+1/3)$,
$\Delta t({\rm JAVELIN};\ \alpha_{\nu}=0)$, and
$\Delta t({\rm JAVELIN};\ \alpha_{\nu}=+1/3)$, respectively.}
\end{deluxetable}

\end{document}